\documentclass[]{aa}  
\usepackage[T1]{fontenc}
\pdfoutput=1 
\usepackage{ae,aecompl}
\usepackage{amssymb,amsmath,amsfonts}
\usepackage{graphicx}
\graphicspath{{./images/}}
\usepackage{txfonts}
\usepackage{hyperref}
\hypersetup{colorlinks=true,linkcolor=blue,citecolor=blue,filecolor=blue,urlcolor=blue}
\usepackage{subfig}
\usepackage{mathptmx}

\newcommand{\degree}{\hbox{$^{\circ}$}}
\renewcommand{\arcsec}{\hbox{\,arcsec}}
\newcommand{\sqdeg}{deg$^2$}

\newcommand{\ddfacet}{\textsc{DDFacet}}
\newcommand{\wsclean}{\textsc{WSClean}}
\newcommand{\aoflagger}{\textsc{aoflagger}}
\newcommand{\dppp}{\textsc{dppp}}
\newcommand{\losoto}{\textsc{losoto}}
\newcommand{\pybdsf}{\textsc{PyBDSF}}

\newcommand{\muJybeam}{\hbox{$\mu$Jy\,beam$^{-1}$}}  
\newcommand{\mJy}{\hbox{mJy}}
\newcommand{\mJybeam}{\hbox{mJy\,beam$^{-1}$}}

\usepackage[normalem]{ulem}
\usepackage{color}

\begin{document} 

\defcitealias{ScaifeHeald2012}{SH12}

   \title{The LOFAR LBA Sky Survey: Deep Fields  I. The Bo\"otes Field\thanks{The image, full catalogue (Table\,\ref{tab:cat}), and matched catalogue (Section\,\ref{sect:deepmatch}) are only available in electronic form at the CDS via anonymous ftp to \href{cdsarc.u-strasbg.fr} (\href{ftp://130.79.128.5}{130.79.128.5}) or via \url{http://cdsweb.u-strasbg.fr/cgi-bin/qcat?J/A+A/}.}}  
   \titlerunning{LoLSS-Deep Bo\"otes}

   \author{W.\,L.~Williams\inst{1} \and
           F.~de~Gasperin\inst{2} \and
           M.~J.~H.~Hardcastle\inst{3} \and
           R.~van~Weeren\inst{1} \and
           C.~Tasse\inst{4,5}
           T.~W.~Shimwell\inst{6,1}
           P.~N.~Best\inst{7} \and
           M.~Bonato\inst{8,9,10} \and
           M.~Bondi\inst{8} \and
           M.~Br\"uggen\inst{2} \and
           H.\,J.\,A.~R\"ottgering\inst{1} \and
           D.~J.~B.~Smith\inst{3} 
           }
   \authorrunning{Williams et~al.}
   \institute{Leiden Observatory, Leiden University, PO Box 9513, NL-2300 RA Leiden, the Netherlands\\ 
              \email{wwilliams@strw.leidenuniv.nl}
              \and
              Hamburger Sternwarte, Universit\"at Hamburg, Gojenbergsweg 112, 21029, Hamburg, Germany 
              \and
              Centre for Astrophysics Research, Department of Physics, Astronomy and Mathematics, University of Hertfordshire, College Lane, Hatfield AL10 9AB, UK 
              \and 
              GEPI and USN, Observatoire de Paris, CNRS, Universit\'e Paris Diderot, 5 place Jules Janssen, 92190 Meudon, France  
              \and 
              Centre for Radio Astronomy Techniques and Technologies, Department of Physics and Electronics, Rhodes University, Grahamstown 6140, South Africa 
              \and
              ASTRON, Netherlands Institute for Radio Astronomy, Oude Hoogeveensedijk 4, 7991 PD, Dwingeloo, The Netherlands 
              \and
              Institute for Astronomy, University of Edinburgh, Royal Observatory, Blackford Hill, Edinburgh, EH9 3HJ, UK 
              \and
              INAF-IRA, Via Gobetti 101, I-40129, Bologna, Italy 
              \and
              Italian ALMA Regional Centre, Via Gobetti 101, I-40129, Bologna, Italy 
              \and
              INAF-Osservatorio Astronomico di Padova, Vicolo dell'Osservatorio 5, I-35122, Padova, Italy
              }

   \date{Received 8 July 2021 / Accepted 16 September 2021}

\abstract{We present the first sub-mJy ($\approx0.7$\,{\mJybeam}) survey to be completed below $100$\,MHz, which is over an order of magnitude deeper than previously achieved for widefield imaging of any field at these low frequencies. The high-resolution ($15 \times 15${\arcsec}) image of the Bo\"otes field at $34$--$75$\,MHz is made from $56$ hours of observation with the LOw Frequency ARray (LOFAR) Low Band Antenna (LBA) system. The observations and data reduction, including direction-dependent calibration, are described here. We present a radio source catalogue containing $1,948$ sources detected over an area of $23.6$\,\sqdeg{},  with a peak flux density threshold of $5\sigma$.  Using existing datasets, we characterise the astrometric and flux density uncertainties, finding a positional uncertainty of $\sim1.2$\,\arcsec{}  and a flux density scale uncertainty of about $5$~per~cent. Using the available deep $144$-MHz data, we identified $144$-MHz counterparts to all the $54$-MHz sources, and produced a  matched catalogue within the deep optical coverage area containing $829$ sources. We calculate the Euclidean-normalised differential source counts and investigate the low-frequency radio source spectral indices between $54$ and $144$\,MHz. Both show a general flattening in the radio spectral indices for lower flux density sources, from $\sim-0.75$ at 144-MHz flux densities between $100$ and $1000$\,mJy to $\sim-0.5$ at  144-MHz flux densities between $5$ and $10$\,mJy.  Such flattening is attributable to a growing population of star forming galaxies and compact core-dominated AGN.}


   \maketitle
%
\section{Introduction}


The LOw Frequency ARray \citep[LOFAR;][]{vanHaarlem2013} has been operating successfully for a number of years now and has already produced a number of interesting results, including the first releases of the full sky survey \citep{Shimwell2017,Shimwell2019}, and of a series of Deep Fields \citep{Tasse2021,Sabater2021} produced by the Surveys Key Science Project (KSP). However, most of the imaging work to date has focused on the upper end of LOFAR's frequency coverage, using the High Band Antenna (HBA) system, operating at 120--240\,MHz. The greater challenges in imaging at lower frequencies have meant that, until recently,  less work has focused on imaging with the Low Band Antenna (LBA) system at $10$--$75$\,MHz. The lower sensitivity of the LBA dipoles and the wide field of view (FoV; $\sim20$\,\sqdeg{}) of the LBA stations both compound the problem of correcting for the propagation errors introduced by the varying ionosphere above the array \citep[e.g.][]{Mevius2016,deGasperin2018b}. The magnetised plasma of the ionosphere causes, to first order, propagation  delays (hence phase errors) inversely proportional to frequency ($1/\nu$), which vary on temporal scales of tens of seconds and spatial scales  of a few arcminutes. This therefore requires several tens of different corrections to be determined and applied within the FoV of a single observation. Higher order effects, including Faraday rotation ($\propto 1/\nu^2$), are also non-negligible at these frequencies.
 
Early LBA work was limited to very bright sources, such as M87 \citep{deGasperin2012} and 3C\,295 \citep{vanWeeren2014}, and wider field imaging such as that of the Bo\"otes field \citep[$\sim5$\,\mJybeam{} noise;][]{vanWeeren2014} did not have full ionospheric calibration. In contrast, HBA observations are now routinely processed to produce thermal-noise images with a pipeline \citep[\textsc{ddf-pipeline};][]{Shimwell2019,Tasse2021} that was built using \textsc{killMS} \citep{Tasse2014a,Tasse2014b,SmirnovTasse2015} for {direction-dependent} calibration and \ddfacet{} \citep{Tasse2018} for wide-field correction and imaging.
Only recently have we successfully implemented these and other {direction-dependent} calibration and imaging algorithms to work at the lower frequencies of the LBA, leveraging advances in other software, including  \dppp{} \citep{vanDiepen2018} and \wsclean{} \citep{Offringa2014}.  Together with a full understanding of the systematic effects in LOFAR data \citep{deGasperin2018b,deGasperin2019}, the application of {direction-dependent} calibration has led to a number of high-resolution ($\sim15$\,\arcsec{}), low-noise ($\sim1$\,\mJybeam{}) images of several individual targets, including the Toothbrush cluster \citep{deGasperin2020a}, Abell\,1758 \citep{Botteon2020} and the planetary system HD\,80606 \citep{deGasperin2020b}.
This has also allowed for the start of the first LOFAR low-frequency wide-area surveys with the LOFAR LBA Survey \citep[LoLSS;][]{deGasperin2021}. Here we apply a modified version of the \textsc{ddf-pipeline} to significantly longer LBA observations than those published so far. This provides a crucial test of the expectation that the noise continues to decrease with integration time, and is not fundamentally limited by systematic effects.

Similar to the tiered approach being undertaken for surveys with the HBA, the LOFAR Two-metre Sky Survey \citep[LoTSS;][]{Shimwell2019}, and its Deep Fields \citep{Tasse2021,Sabater2021}, we aim to complement LoLSS with several deep LBA observations of the LoTSS Deep fields (Bo\"otes, ELAIS-N1 and Lockman Hole). This is the first such observation. This deep, very low-frequency radio imaging, in combination with the higher frequency LoTSS Deep fields, will enable low-frequency spectral indices to be determined for all sources detected at LBA and reveal rare, extreme-valued sources. This will provide unique data on the shape of the low-frequency spectra of star forming galaxies, AGN, and galaxy clusters that will help answer many questions related to these types of sources. LOFAR LBA surveys reach an order of magnitude deeper and achieve an order of magnitude higher resolution than previous surveys at these frequencies \citep[VLSSr;][]{Lane2014}.

As one of several extragalactic deep fields spanning a few square degrees, there is a  wealth of additional multi-wavelength data available for the Bo\"otes field. This field was originally part of the National Optical Astronomy Observatory (NOAO) Deep Wide Field Survey \citep[NDWFS;][]{Jannuzi1999}, which covered $\sim9$\,\sqdeg{} in the optical and near-infrared $B_W$, $R$, $I,$ and $K$ bands. Further observations have since been obtained, including X-ray \citep{Murray2005,Kenter2005}, UV \citep[GALEX;][]{Martin2003}, and mid-infrared \citep{Eisenhardt2004,Martin2003} imaging. Moreover, the AGN and Galaxy Evolution  Survey (AGES) has provided redshifts for 23,745 sources, including both normal galaxies and AGN, across 7.7\,\sqdeg{} of the field \citep{Kochanek2012}. As one of the LoTSS Deep Fields, high-quality photometric redshifts have been determined for over 2 million optical sources \citep{Duncan2021}.  The Bo\"otes field has been widely surveyed at radio wavelengths --  with the WSRT at 1.4\,GHz \citep{deVries2002} and the VLA at 1.4\,GHz \citep{Higdon2005} and 325\,MHz \citep{Croft2008,Coppejans2015}. Additionally, Bo\"otes is one of the LoTSS Deep Fields, with a sensitivity achieved thus far of 30\,\muJybeam{} using the HBA at $144$\,MHz \citep{Tasse2021}. Finally, the previous LOFAR LBA image of the  Bo\"otes field by \cite{vanWeeren2014} at $54$--$70$\,MHz covered $19.4$\,\sqdeg{} and reached a noise level of $4.8$\,{\mJybeam} with a beam size of $31 \times 19${\arcsec}.

The outline of this paper is as follows. In Section~\ref{sect:obs} we describe the LOFAR LBA observations covering the  Bo\"otes field. In Section~\ref{sect:datared} we describe the data reduction techniques employed to achieve the deepest possible image, with a focus on the development and execution of a robust and automated pipeline for the direction-dependent calibration and imaging.  In Section~\ref{sect:cat} we present the final image and describe the source-detection method and the compilation of the source catalogue. This section also includes an analysis of the quality of the catalogue. The spectral index distribution and differential source counts  are presented in Section \ref{sect:results}. Finally, Section \ref{sect:concl} summarises and concludes this work.

Throughout this paper, the spectral index, $\alpha$, is defined as $S_{\nu} \propto \nu^\alpha$, where $S$ is the source flux density and $\nu$ is the observing frequency.  

\section{Observations}
\label{sect:obs}

\begin{figure}
 \centering
 \includegraphics[width=0.495\textwidth]{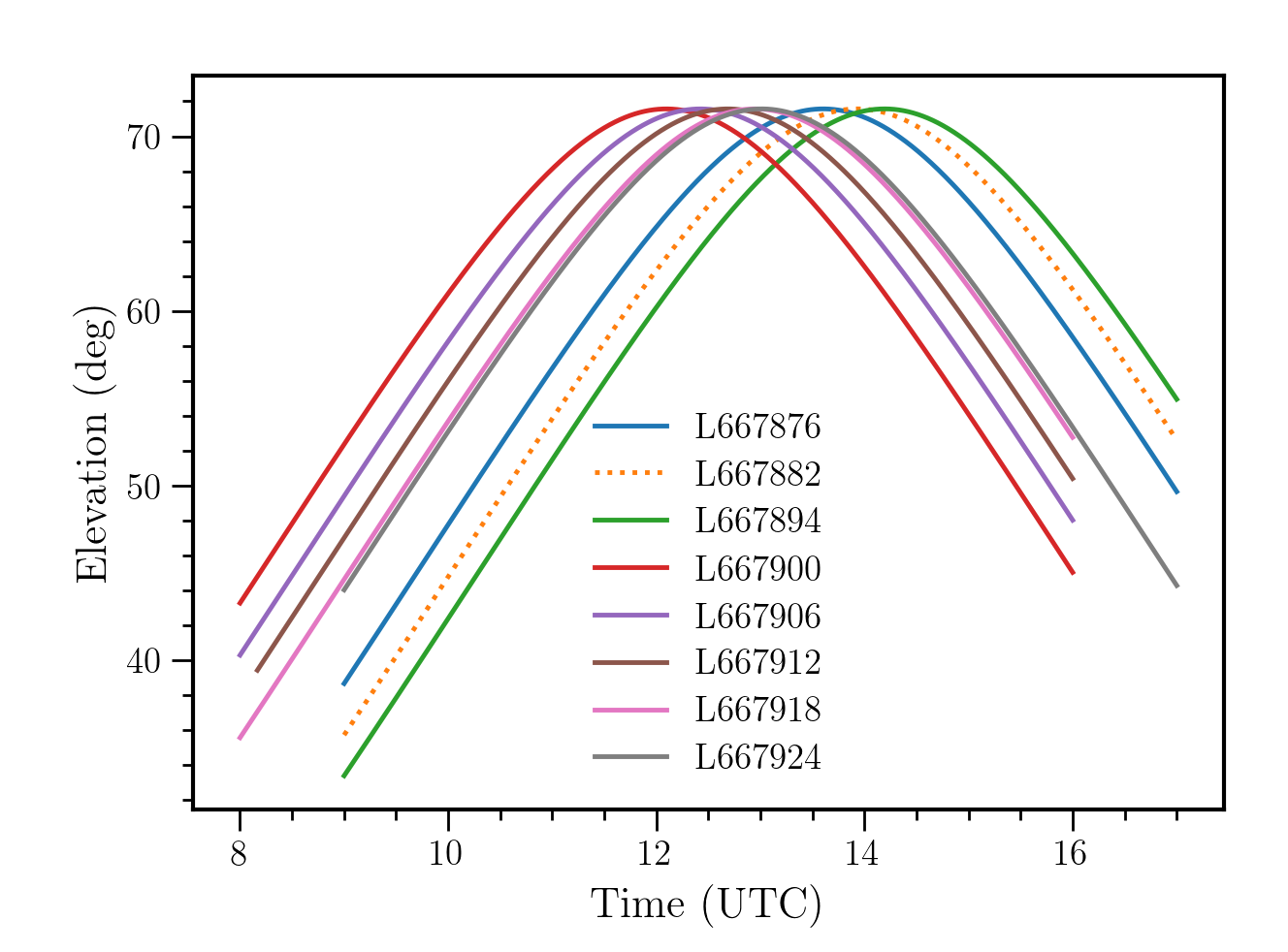}
\caption{Elevation plots for the eight observations.}
\label{fig:elevation}
\end{figure}

The Bo\"{o}tes field, at 14h32m03.0s +34d16m33s, was observed during September--October 2018 with the LOFAR Low Band Antenna (LBA) stations for a total of $64$\ hr spread over eight observations of 8 hr each under 
Project code LC10\_007.  A summary of the observations is given in Table~\ref{tab:observations}. Each eight-hour track was roughly centred at transit, with the elevation of the target field $\gtrsim35$\,deg (see Fig~\ref{fig:elevation}). All four correlation products (XX, XY, YX and YY) were recorded with the frequency band divided into $195.3125$\,kHz-wide sub-bands (SBs) with each SB further divided into $64$ channels. The integration time used was $1$\,s. These high time and frequency resolutions were selected to allow for the accurate removal of radio frequency interference (RFI). The maximum number of SBs for the system in $8$-bit mode is $488$ and the chosen strategy was to use $244$ for the Bo\"otes field giving a total bandwidth of $48$\,MHz between $30$ and $78$\,MHz. The remaining $244$ SBs were used to observe the bright calibrator source,  3C\,295, located $18.3${\degree} away,  with a simultaneous station beam and an identical SB setup to the Bo\"otes observation.

\begin{figure}
 \centering
 \includegraphics[width=0.499\textwidth]{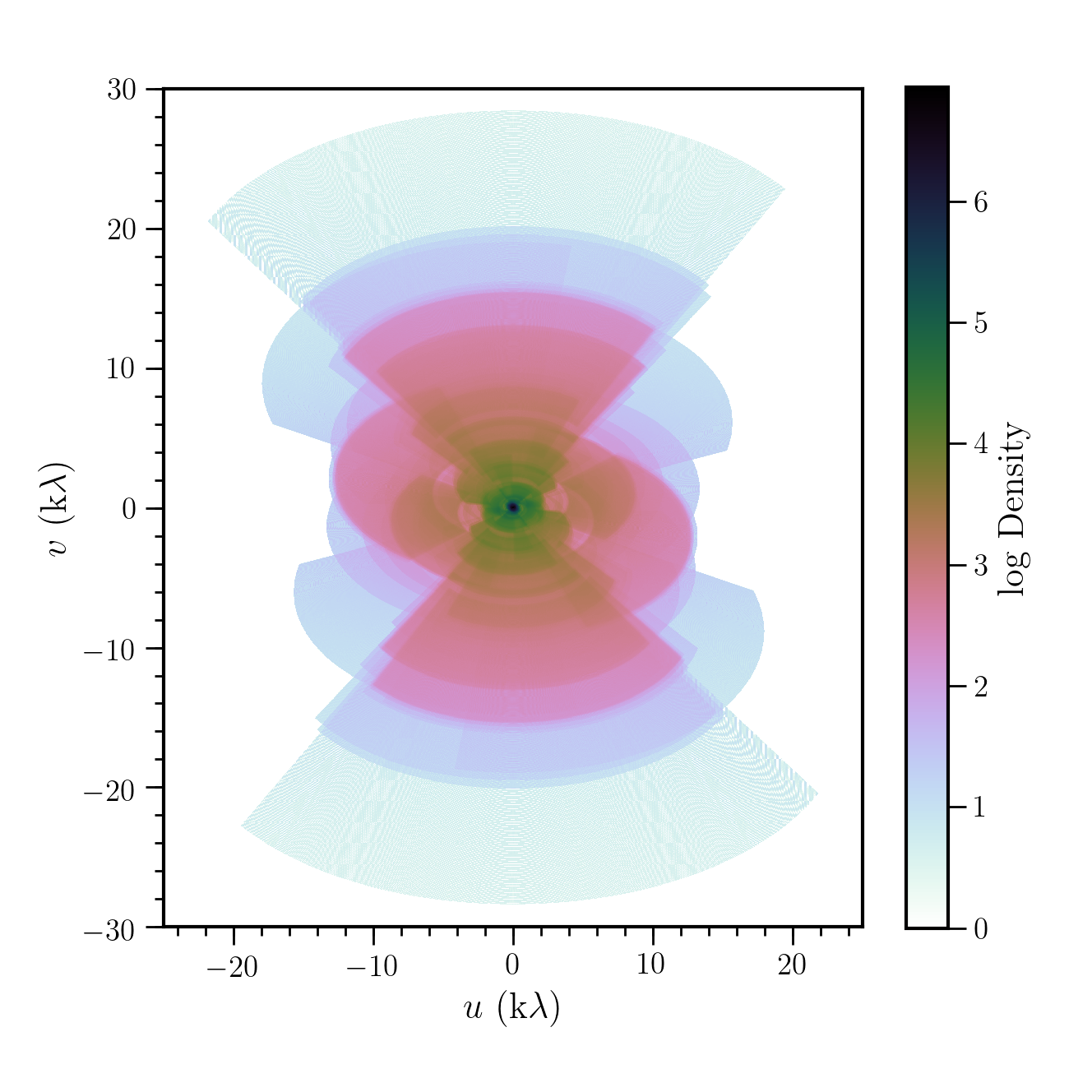}
\caption{$uv$-coverage for one observation of the Bo\"otes field. The maximum baseline is $120$\,km (or $30$\,k$\lambda$). The colour scale shows the logarithm of the density of points in $uv$-space.}
\label{fig:uvcover}
\end{figure}

The full Dutch array with $14$  remote and $24$ core stations was used for all observations, with between $34$ and $38$ total stations available for each observation (typical station failure rates). This setup results in baselines that range between $40$\,m and $120$\,km. The $uv$-coverage for one Bo\"{o}tes field observation is plotted in Fig.~\ref{fig:uvcover} -- showing the usual dense inner coverage and outer extensions to the north and south due to the high density of core stations and the  north--south extension of the Dutch remote stations. The maximum theoretical resolution achievable with such an array is $10${\arcsec} at 60\,MHz (varying between $20$ and $8${\arcsec} between the lower and upper ends of the frequency band).  Finally, the `LBA\_OUTER' configuration was employed, which uses the outer $48$ out of the $96$ LBA antennae in each station, giving a station diameter of $81$\,m and a full width at half maximum (FWHM) of the field of view $\approx4.3${\degree} at $54$\,MHz (but varying from $\approx7.8$ to $\approx3.1${\degree} between $30$ and $75$\,MHz).

\begin{table}
\caption{Observations of the Bo\"otes field}
\label{tab:observations}
\begin{small}
\begin{center}
 \begin{tabular}{llll}
 \hline
ID & Start of Observation & Stations\tablefootmark{a} &  Image noise \\
                &      &          & (\mJybeam{}) \\
 \hline
L667876 &  28-Sep-2018/09:00 &  38 & 1.72 \\
L667882\tablefootmark{b} &  23-Sep-2018/09:00 &  34 & 2.48 \\
L667894 &  19-Sep-2018/09:00 &  35 & 1.79 \\
L667900 &  21-Oct-2018/08:00 &  37 & 1.82 \\
L667906 &  16-Oct-2018/08:00 &  37 & 1.79 \\
L667912 &  12-Oct-2018/08:10 &  35 & 1.72 \\
L667918 &  08-Oct-2018/08:00 &  37 & 1.69 \\
L667924 &  07-Oct-2018/09:00 &  37 & 1.67 \\
 \hline
 \end{tabular} \\
\end{center}
\tablefoot{\\
\tablefootmark{a}{Including stations CS013 and CS031 that were subsequently flagged for all observations.}\\
\tablefootmark{b}{L667882 was very poor in image quality so we excluded it in the final imaging.}
}
\end{small}
\end{table}

\section{Data reduction}
\label{sect:datared}

\subsection{Pre-processing}
\label{ssect:preproc}
Initial pipeline pre-processing per SB for both the calibrator and target was carried out by the Netherlands Institute for Radio Astronomy (ASTRON) and included `demixing'  \citep[a computationally fast method of removing bright `A-team' sources by subtracting their contributions directly from the visibility data;][]{vanderTol2007} of Cassiopeia A at a distance of $79^\circ$ and Cygnus A at $63^\circ$, flagging of RFI with \aoflagger{} \citep{Offringa2010,Offringa2012}, and final averaging to $4$\,s and $48.828$\,kHz per channel. The final pre-processed data were stored in the LOFAR Long Term Archive \citep[LTA;][]{Belikov2011}. \textsc{Dysco} \citep[][]{Offringa2016} compression was used resulting in a total data size per observation for both the target and calibrator of $\sim300$\,GB.

\subsection{Direction-independent calibration}
\label{sect:dical}
After download from the archive, the data were processed using \textsc{prefactor version 3 (v3)}\footnote{\url{https://github.com/lofar-astron/prefactor/}}  to perform further flagging, calibration, and averaging. The \textsc{prefactor} pipeline is built within the LOFAR  \textsc{genericpipeline} framework to describe the workflow, and relies heavily on \textsc{casacore} tables and measurement sets \citep[MSs;][]{vanDiepen2015}  and the Default Pre-Processing Pipeline \citep[\dppp{};][]{vanDiepen2018} for individual operations.  The full details of the method for calibrating direction-independent effects in LOFAR data ---particularly LBA data--- is described in detail by 
\cite{deGasperin2019,deGasperin2020a} and implemented in \textsc{prefactor v3}. Here, we summarise the steps, noting particular steps where the reduction diverged from the default HBA calibration, which is documented in detail elsewhere \citep[e.g.][]{Shimwell2019}.

\subsubsection{Calibrator processing}
For all SBs of the calibrator beam, the first step was to flag any {data from
known bad stations (CS013 and CS031 in all observations), data at low elevations ($< 20$ deg) and data with extremely low amplitudes ($< 10^{-30}$)}. \aoflagger{} was then run on a virtually concatenated  MS of all SBs to remove RFI across the full bandwidth using the \verb|LBAdefaultwideband| strategy. No additional averaging was done so the calibrator data remain at a resolution of $4$\,s and $48.82$\,kHz per channel ($4$ channels per SB).

The model used for calibration of 3C\,295 consisted of two point-source components separated by $3.93${\arcsec} and was normalised to  the flux scale of \citet[][hereafter SH12]{ScaifeHeald2012}. The visibilities of this model are predicted with \dppp{} and stored in the \textsc{MODEL\_DATA} column of the MS for efficiency.

The following steps of the calibrator pipeline were then to solve sequentially for the following effects:
\begin{enumerate}
\item polarisation alignment (PA) -- a time-independent frequency-dependent ($\propto \nu$) term that corrects for the misalignment between the XX and YY polarisations due to different station calibration for the independent feeds,
\item  Faraday rotation (FR) -- a direction-, time-, and frequency-dependent ($\propto \nu^{-2}$) term that corrects for the phase rotation caused by the ionosphere,
\item bandpass (B) -- a  {time-independent,} frequency-dependent term that, for the LBA, largely corrects for the dipole response{ \citep[strongly peaked at $\sim60$\,MHz, cf. fig.\,20 of][]{vanHaarlem2013}},
\item and phase (P) -- a time- and frequency-dependent scalar term that incorporates both the effects of the ionosphere {($\propto\nu^{-1}$ to first order\footnote{The second-order term is Faraday rotation and the third-order term $\propto\nu^{-3}$ is usually ignored but can become relevant below $40$\,MHz \citep{deGasperin2018b}}}) and instrumental clock errors {($\propto\nu$)}.
\end{enumerate}
At each point, the previously determined correction(s) were applied before solving for the next term. Gain solutions were derived with \dppp{} \textsc{ddecal} to determine the diagonal and rotation terms for the polarisation alignment and Faraday rotation steps, and diagonal-only terms for the bandpass and ionosphere steps at the full frequency and time resolution. To improve the signal-to-noise ratio, at each calibration step initially the  \textsc{DATA} (the pre-processed visibilities), and subsequently \textsc{CORRECTED\_DATA} (the pre-processed visibilities with the previously determined correction(s) applied) were first smoothed with a Gaussian kernel both in time and frequency in a baseline-dependent way before solving, that is the shorter baselines were smoothed with larger kernels than the longer baselines. The corrections were derived from the raw gain solutions with \losoto{}. The final P solutions are decomposed into the contributing ionosphere and clock terms, making use of their different frequency dependence ($\propto 1/\nu$ and $\propto \nu$ respectively). Although these terms are not applied separately to the target visibilities, they are effectively corrected for by applying the time- and frequency-dependent phase solutions. These terms also provide a useful diagnostic as to the severity of the ionospheric effects during each observation.

\subsubsection{Target calibration}
In the same way as for the calibrator data, for all SBs of the target beam, the first step of the pipeline was to flag any
{data from known bad stations (CS013 and CS031 in all observations), data at low elevations ($< 20$ deg), and data with extremely low amplitudes ($< 10^{-30}$)}. The calibration solutions from the corresponding simultaneous calibrator beam were then applied to each target observation, in the following order: PA, B, and P. {We did not apply the FR because this is a direction-dependent effect\footnote{Later versions of the \textsc{LiLF} pipeline, developed after this data was processed, enable Faraday rotation to be corrected for the target field.}.} We note that, unlike the procedure for HBA processing, because the calibrator is observed simultaneously, we correct the target field with the phases from the calibrator. While these contain the instrumental clock delays that are independent of direction, they also contain the ionospheric delays in the direction of the calibrator. Any further ionospheric phase solutions calculated for the target field will therefore be differential with respect to the direction of 3C\,295; however, the frequency-dependence will remain the same. The data were then corrected for the station beams. 
No additional averaging was done so the data remain at a resolution of $4$\,s and $48.82$\,kHz per channel ($4$ channels per SB).

HBA processing usually follows this stage with a step to predict the visibilities from bright A-team sources and flag periods with high contributions. However, as the LBA data had already been demixed in pre-processing, this was not necessary.  The data were then concatenated into bands of $2$\,MHz, and \aoflagger{} was run on a virtually concatenated MS of all bands to remove any remaining RFI, again using the \verb|LBAdefaultwideband| strategy across the full bandwidth.

We did not use \textsc{prefactor} for the final calibration of the target field, but rather used the Libraries for Low Frequency (\textsc{LiLF}) framework\footnote{\url{https://github.com/revoltek/LiLF}} for its ease of testing different strategies.
We used \texttt{BLSmooth} to do a baseline-based smoothing of the visibilities before solving for scalarphase only  and used a model for the field from the Global Sky Model\footnote{\url{https://lcs165.lofar.eu/}} that included information from TGSS, WENSS, NVSS, and VLSS. We included sources predicted to be above $0.5$\,Jy at $60$\,MHz  within a radius of $0.67$ times the FWHM of the station beam at $34$\,MHz. Phases were solved for on a timescale of $12$\,s and within frequency blocks of $0.5$\,MHz. We note that this is the highest time resolution used in the data reduction so any errors still present at these timescales cannot be removed later in the  direction-dependent calibration.

\begin{figure*}
 \centering
\includegraphics[width=0.9\textwidth]{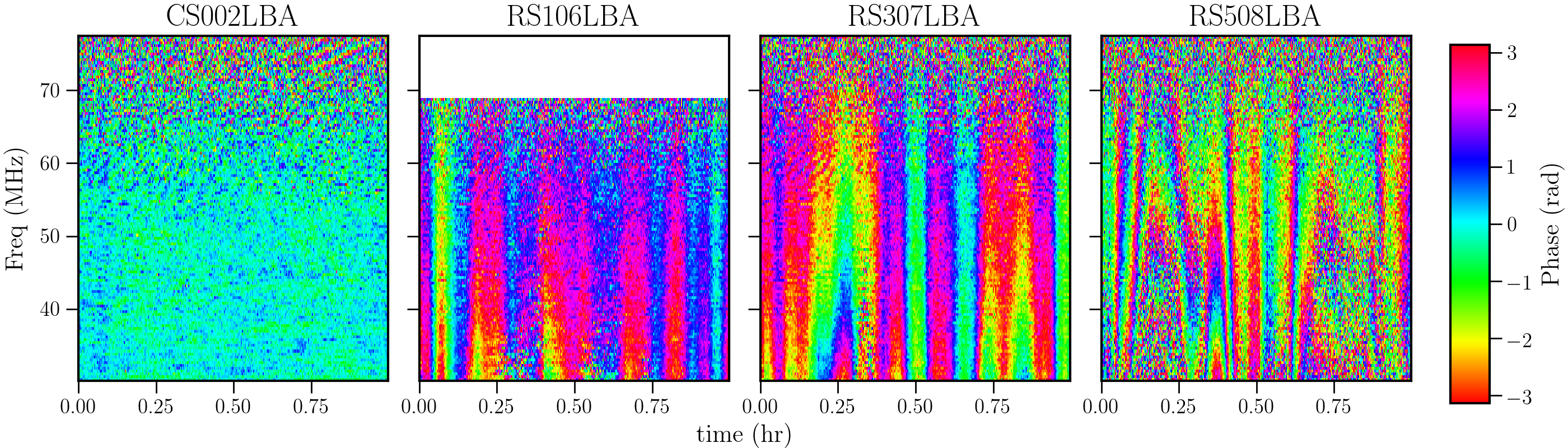}\\
\includegraphics[width=0.9\textwidth]{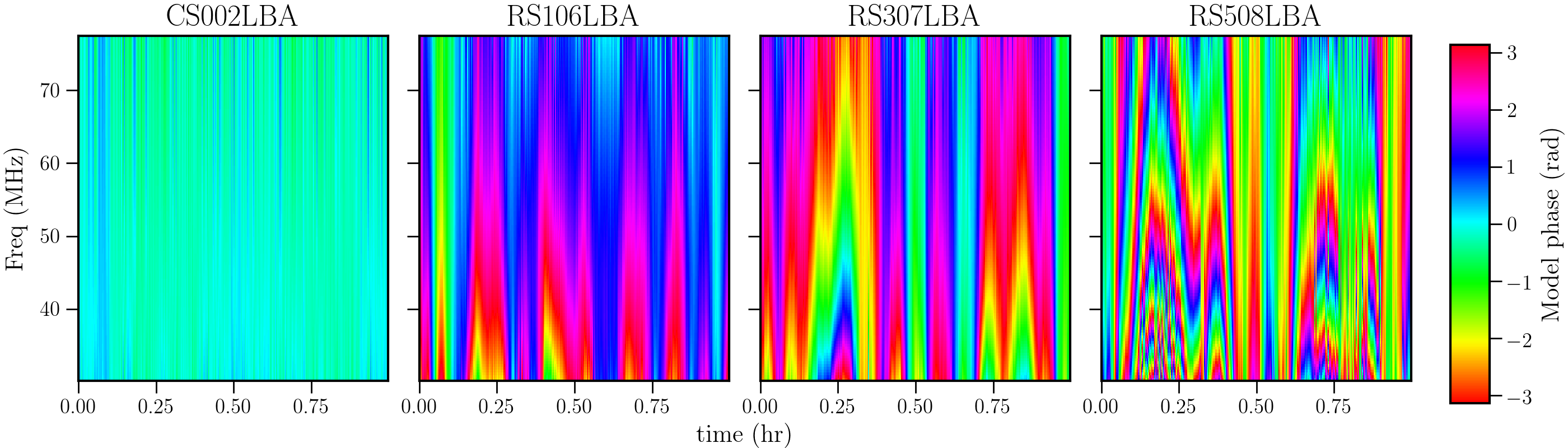}\\
\includegraphics[width=0.9\textwidth]{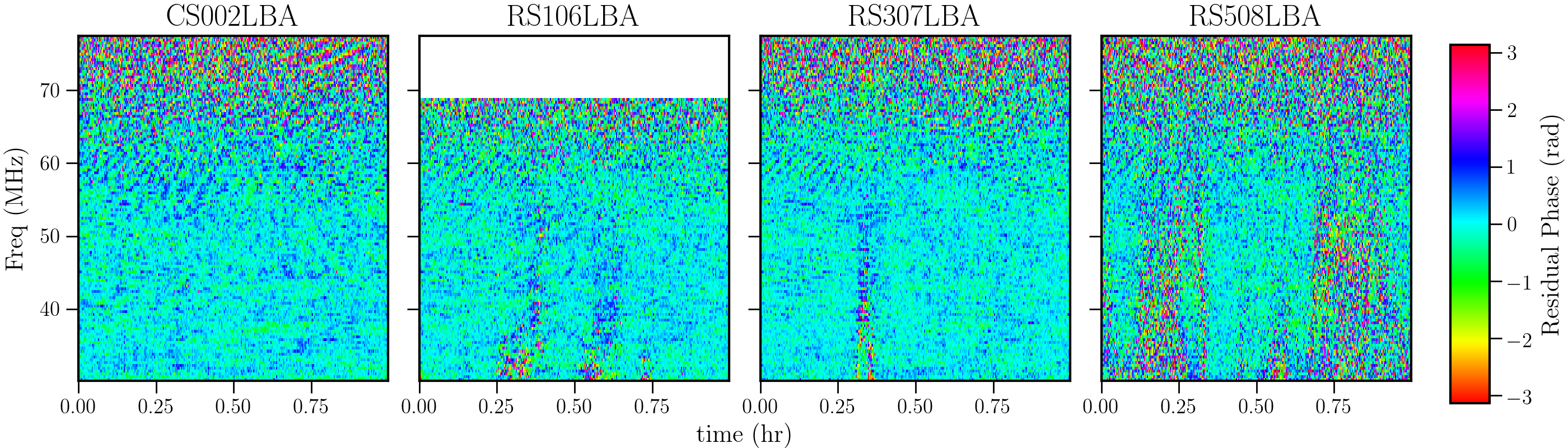}\\
\includegraphics[width=0.9\textwidth]{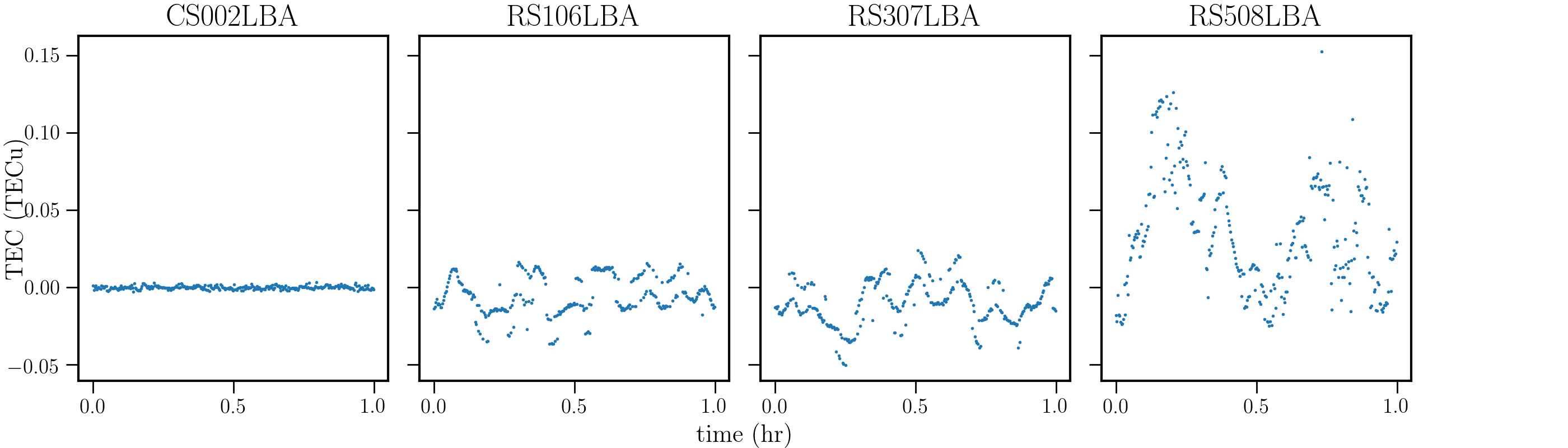}\\
\includegraphics[width=0.9\textwidth]{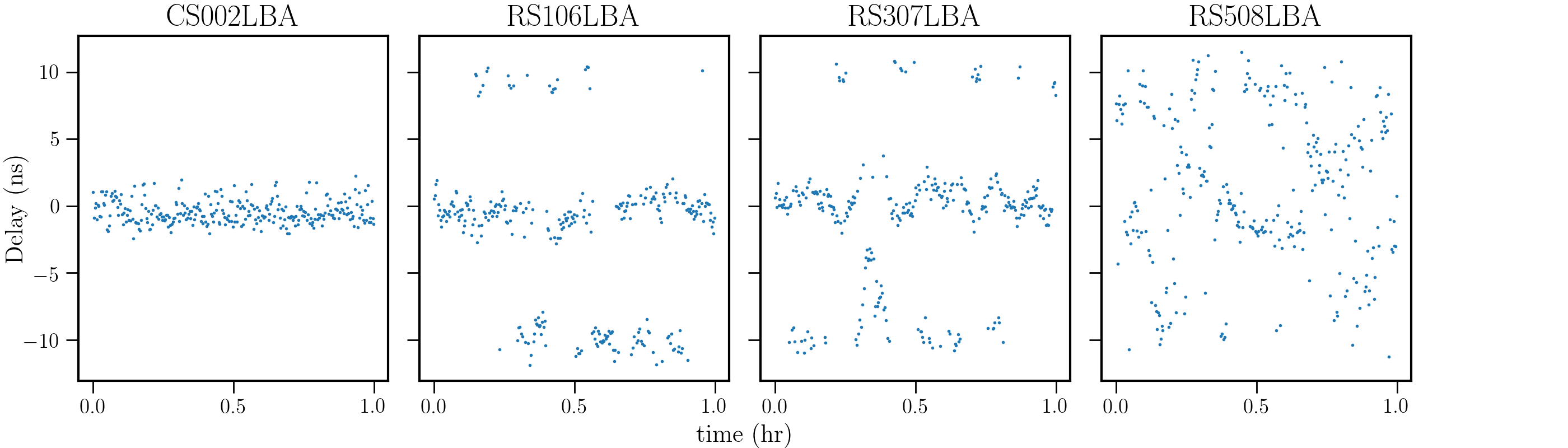}\\
\caption{{Example direction-independent phase solutions on the target field, taken from the fourth hour of observation L667924.  From top to bottom the panels show the raw phase solutions, the TEC+Delay model phases, the phase residuals, and the fitted TEC and Delay terms as a function of time.}}
\label{fig:eg_fitsols}
\end{figure*}

Experience has shown that a single {total electron content (TEC)} term is insufficient to characterise the phase solutions, and so we decompose the scalar phases into a TEC ($\propto 1/\nu$) term and a delay ($\propto \nu$) term. The origin of this delay term is not fully understood because the instrumental clock errors are corrected when the phases are transferred from the calibrator. It may be a geometric delay due to pointing errors. An example of the scalar phase solutions and fitted TEC and delay terms is shown in Fig.\,\ref{fig:eg_fitsols} for a single one-hour chunk of data from one observation{ that is representative of the general quality of the solutions. We note that, particularly for remote stations, the delay solutions during some time periods (most often at low elevation) can be very noisy, even random. This may be due to improperly modelled phases (Faraday rotation and higher order ionospheric effects are not modelled here), poor signal-to-noise ratio, or strongly direction-dependent effects that mean a single correction for the field is insufficient.} While the noisy data result in discrete `jumps' in the TEC and Delay solutions, due to the incorrect local minimum found in the fit, the overall residual phases are mostly flat and noise-like{ and are flatter than after a TEC-only fit}.
Once these solutions had been determined, the unsmoothed visibility data were then corrected for the combined TEC and delay.

\subsection{Direction-dependent calibration}
\label{sect:ddcal}

The data were further processed using a slightly modified version of  version 2.2 of the standard Surveys KSP pipeline, \textsc{ddf-pipeline}\footnote{\url{https://github.com/mhardcastle/ddf-pipeline}} as described by \cite{Shimwell2019} and \cite{Tasse2021}. This pipeline carries out direction-dependent calibration using \textsc{killMS}\footnote{\url{https://github.com/saopicc/killMS}} \citep{Tasse2014a,Tasse2014b,SmirnovTasse2015} and imaging is done using \ddfacet{}\footnote{\url{https://github.com/saopicc/DDFacet}} \citep{Tasse2018}. Version 2.2 of the pipeline makes use of enhancements to the
calibration and imaging ---particularly of extended sources--- that were
described briefly in section 5 of \cite{Shimwell2019} and discussed more fully by \cite{Tasse2021}. The implementation and modifications we made are described here in detail.

We first processed each of the eight observations separately. This provides information on the variation of quality between observations. We adopt a very conservative approach and use only five directions with the same facet layout for each of the eight observations. The small number of directions is a trade-off between the facet size, which is large (approximately $4$--$6$\,square degrees), and $S/N$. To process a single observation, we configured  the pipeline to perform two cycles of phase-only self-calibration, with imaging at a lower resolution ($30${\arcsec}), followed by one cycle of phase-only calibration and imaging at $15$\,{\arcsec}. Solutions in both cases are obtained on timescales of $2$\,min and within each $2$-MHz band. In practice, full-Jones solutions are calculated, but the amplitudes are set to unity and only the phases are applied in imaging. The phase component of the solutions is smoothed with a TEC-like function, that is $1/\nu$, over the full bandwidth. A final very slow full-Jones calibration is performed, providing both phase and amplitude solutions on timescales of $43$\,min. Unlike the standard LoTSS-DR2 processing, we do no additional direction-independent calibration; we  apply amplitude corrections only on the very long timescale, and we use the full bandwidth already in the initial steps. As the LBA beam model is better understood than that of the HBA, we did not apply the bootstrapping of the flux-density scale usually done for the HBA.
We only processed bands 2--22 ($34$--$75$\,MHz), avoiding the low-$S/N$ high-frequency bands \citep[where the dipole response rapidly drops off; cf. fig.\,20 of][]{vanHaarlem2013} and the low-frequency end where the ionospheric effects become too large (in particular Faraday rotation). The noise levels achieved in each eight-hour observation, listed in Table\,\ref{tab:observations}, mostly vary between 1.67 and 1.82\,\mJybeam{}, with one strong outlier of 2.48\,\mJybeam{} for L667882. We therefore excluded this observation from further processing. 

To process the seven remaining observations together, we configured the pipeline in the deep-imaging mode \citep[as described in][]{Tasse2021} using the image from the best single observation (L667924) as a
starting model. Each $2$-MHz band is calibrated against this model: first for the $2$\,min phase-only solutions, again smoothed with a TEC-like function, followed by the $43$\,min\footnote{{This is a somewhat arbitrary number, used by default for the HBA pipeline, but captures the slowly-varying amplitudes and phases errors largely caused by the station beam.}} amplitude and phase solutions. Final imaging is done at $15$\arcsec{} resolution. In the final imaging step, we applied a per-facet position offset using this capability in \ddfacet{}. For LoTSS-DR1 and the LoTSS Deep Fields, the per-facet offsets are derived relative to Pan-STARRs \citep{Chambers2016}. However, the resolution of the LBA image is too low and the Pan-STARRS source density too high to achieve unique matches. Instead, the per-facet offsets were derived with respect to the positions of compact sources in the $6$\,\arcsec{} resolution deep HBA image of the Bo\"otes field \citep{Tasse2021} -- compact sources were selected as those with size $<10$\arcsec{}. In summary, after generating a source catalogue using \pybdsf{} and selecting good sources (positional error $<3$\,\arcsec{} and size $<20$\,\arcsec{}), the LBA sources were matched to the HBA compact sources taking the nearest source within $10$\,\arcsec{}, and Gaussians fitted to the position offset distributions. This is implemented within \textsc{ddf-pipeline}. Around $50$ matches within each facet were found, and were of the order of $1$--$3$\,\arcsec{}. These offsets were applied on a per-facet basis in the final round of \ddfacet{} imaging.


\subsection{Final image and catalogue} 
\begin{figure*}
 \centering
\includegraphics[width=\textwidth]{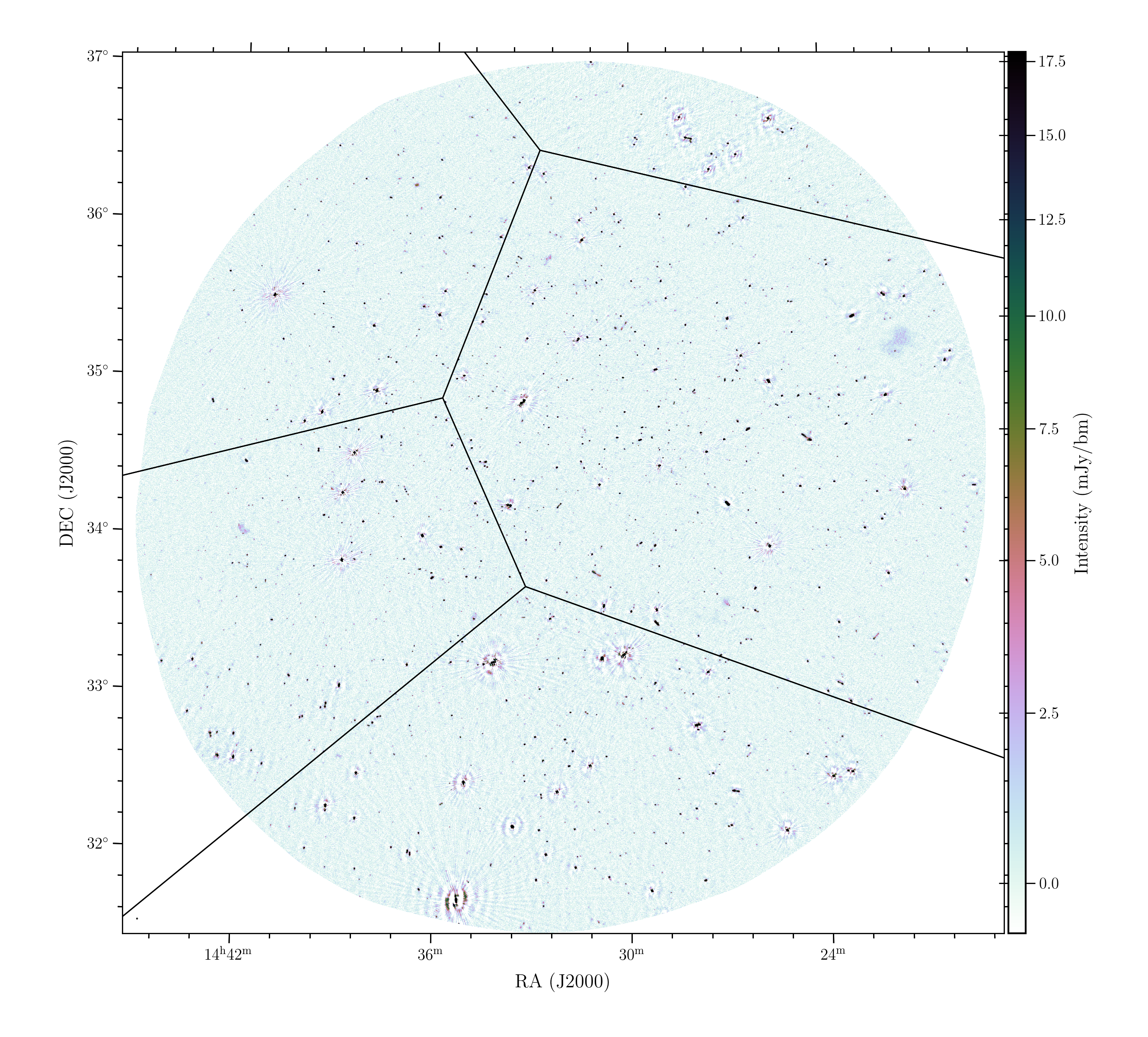}
\caption{Colour-scale map showing the full primary-beam-corrected image. The image covers $23.6$\,\sqdeg{}. The colour scale shows the intensity from $-3\sigma_{\mathrm{cen}}$ to $25\sigma_{\mathrm{cen}}$ where  $\sigma_{\mathrm{cen}} = 0.7$\,{\mJybeam} is the approximate rms in the image centre. The black polygons show the facets used in the calibration and imaging.}
\label{fig:image}
\end{figure*}

\begin{figure*}
 \centering
\includegraphics[width=\textwidth, trim=0.2cm 0.6cm 0.2cm 0.2cm, clip]{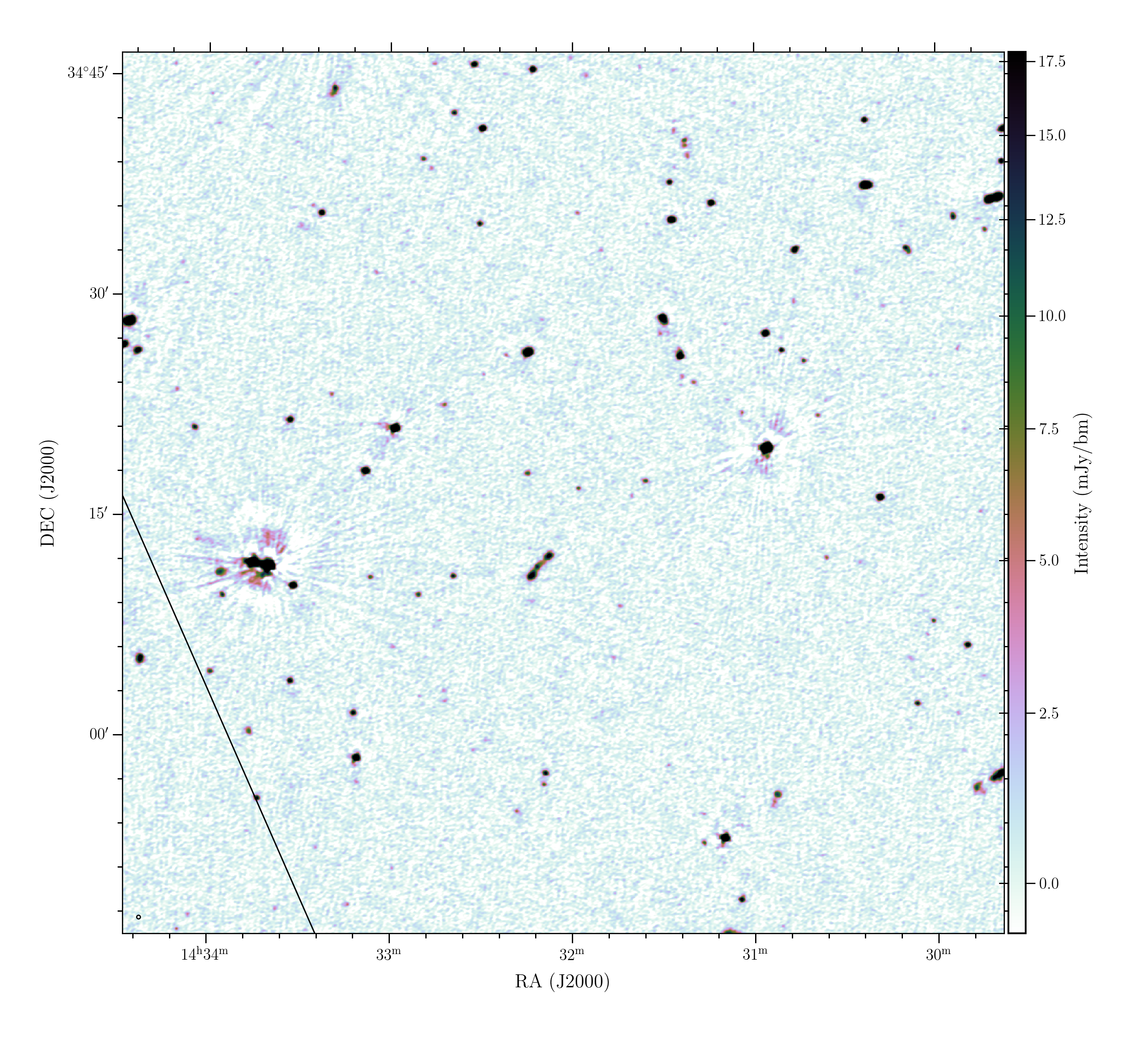}
\caption{Zoom-in of the central part of the primary-beam-corrected image. The image covers $1$\,\sqdeg{}. The colour scale shows the flux density from  $-3\sigma_{\mathrm{cen}}$ to $25\sigma_{\mathrm{cen}}$ where  $\sigma_{\mathrm{cen}} = 0.7$\,{\mJybeam} is the approximate rms in the image centre.  The black line shows one of the facet boundaries used in the calibration and imaging -- even sources on or near the facet boundaries are well imaged.}
\label{fig:image_zoomin}
\end{figure*}

\begin{figure*}
 \centering
\includegraphics[width=0.49\textwidth]{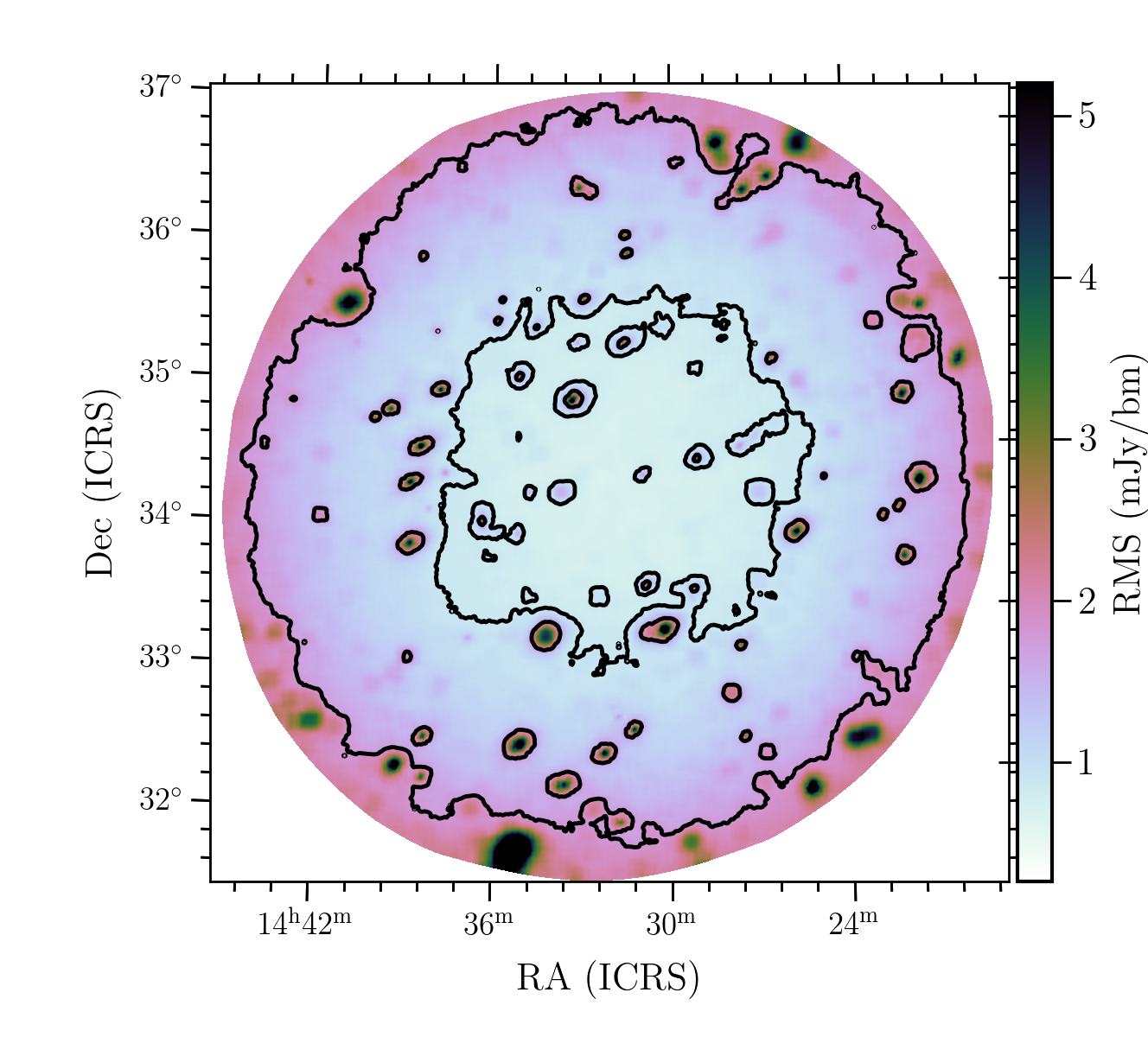}
\includegraphics[width=0.49\textwidth]{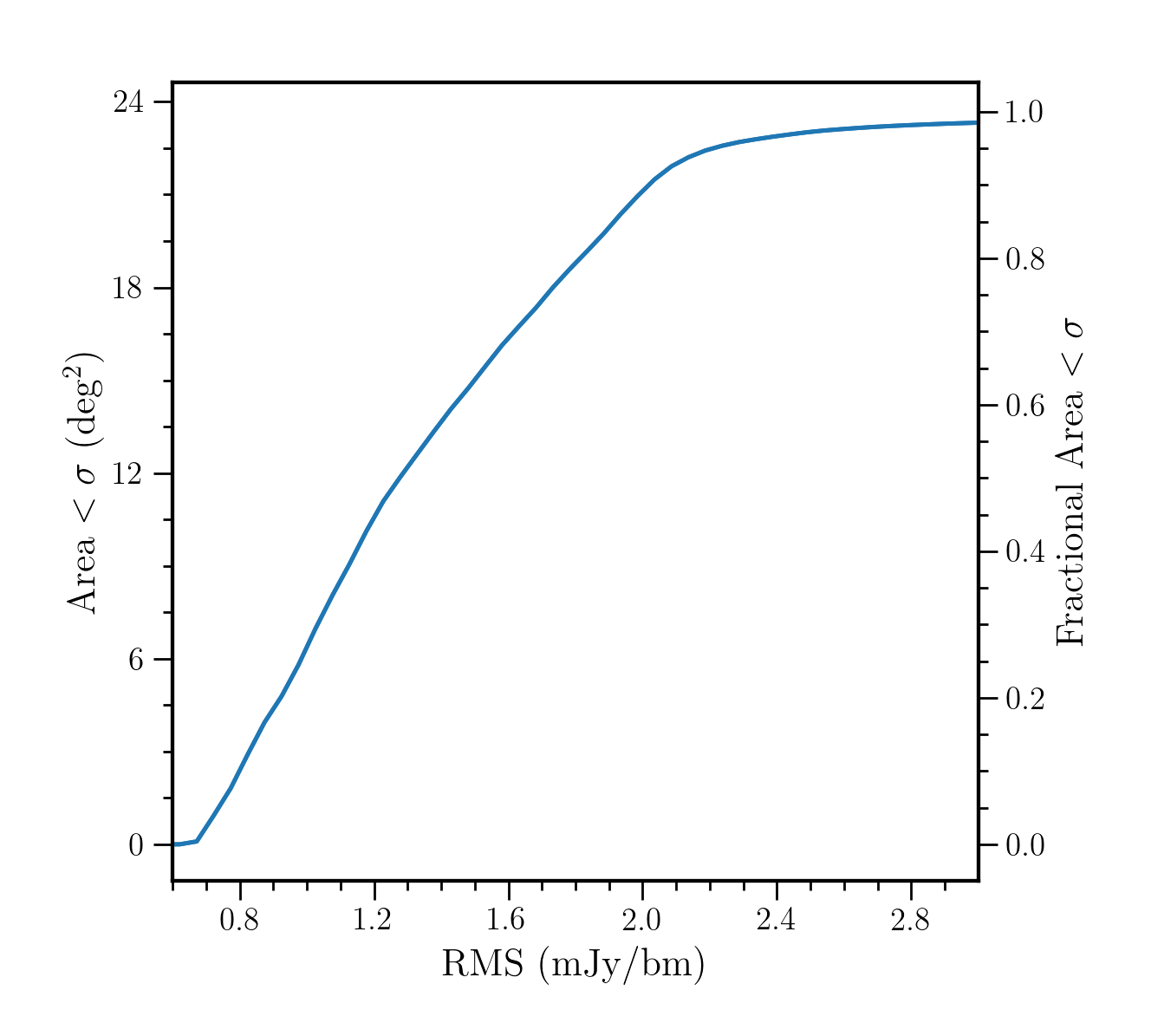}
\caption{Left:  Local rms noise measured in the final image. Contours are plotted at $0.9$\,{\mJybeam} and $1.8$\,{\mJybeam}.  Peaks in the local noise coincide with the locations of bright sources. Right: Cumulative area of the map with a measured rms noise level below the given value.}
\label{fig:mosrms}
\end{figure*}

The final beam-corrected image at $15 \times 15${\arcsec} resolution is shown in  Fig.\,\ref{fig:image} and is masked where the primary beam correction exceeds $3$. This image is available online\footnote{\label{note1}\url{www.lofar-surveys.org}}. A small portion of the image covering the inner $0.25$\,\sqdeg{} is shown in Fig.\,\ref{fig:image_zoomin} to illustrate the resolution and quality of the image. The rms noise level in the central part is relatively smooth with a median value of $\sim0.70$\,{\mJybeam} within the innermost square degree, and $50$\%\ of the map is at a noise level below $1.2$\,{\mJybeam} (see also Fig.\,\ref{fig:mosrms}). There remain some strong phase artefacts {and corresponding increases in noise level} around the brightest sources; these were not entirely removed during the direction-dependent calibration, but are localised. 

\section{Source catalogue}
\label{sect:cat}

To produce a catalogue of the radio sources we extracted sources from the final image using \pybdsf{} \citep{Mohan2015}. This was done using the standard HBA Surveys settings \citep{Shimwell2019}: that is with a peak detection threshold of $5\sigma$ and an island detection threshold of $4\sigma$. The background noise variations were estimated across the images using a sliding box with a box size of $30 \times 30$ synthesised beams, which was decreased to $12 \times 12$ synthesised beams in regions of high $S/N$ sources ($\geq150$) in order to more accurately capture the increased noise level around bright sources. The \pybdsf{} wavelet decomposition was used to  better characterise the complex low-surface-brightness extended emission. Source detection was done on the apparent sky image, while source parameters were extracted from the beam-corrected image. Figure\,\ref{fig:mosrms} shows the variation in rms noise determined across the image. The increase in rms noise towards the edge of the field is a result of the `primary' beam (LOFAR station beam).  Errors on the fitted source shape parameters are computed following \cite{Condon1997}. The \pybdsf{} catalogue contains $1,948$ sources, with total flux densities between $3$\,mJy and $18$\,Jy. The \pybdsf{} source catalogue is available online\textsuperscript{\ref{note1}}. A sample of the catalogue, showing the brightest and faintest entries, is given in Table \ref{tab:cat}.

\begin{sidewaystable*}
\caption{Example entries from the full source catalogue.}
\label{tab:cat}
\begin{small}
\centering
 \begin{tabular}{rccccr@{\,$\pm$\,}lr@{\,$\pm$\,}lccr@{\,$\pm$\,}lr@{\,$\pm$\,}lr@{\,$\pm$\,}l}
 \hline
 \multicolumn{1}{c}{Source Name} & \multicolumn{1}{c}{RA} & \multicolumn{1}{c}{$\sigma_{RA}$} & \multicolumn{1}{c}{DEC}  & \multicolumn{1}{c}{$\sigma_{DEC}$} & \multicolumn{2}{c}{$S_i$}  & \multicolumn{2}{c}{$S_p$} & \multicolumn{1}{c}{$\sigma_{\mathrm{local}}$}  & \multicolumn{1}{c}{$N_{\mathrm{gauss}}$} & \multicolumn{2}{c}{$a$} & \multicolumn{2}{c}{$b$} & \multicolumn{2}{c}{$\phi$} \\
  & \multicolumn{1}{c}{(deg)} & \multicolumn{1}{c}{($\arcsec$)} & \multicolumn{1}{c}{(deg)} & \multicolumn{1}{c}{($\arcsec$)} & \multicolumn{2}{c}{(mJy)} & \multicolumn{2}{c}{(mJy/bm)} & \multicolumn{1}{c}{(mJy/bm)} &  & \multicolumn{2}{c}{($\arcsec$)} & \multicolumn{2}{c}{($\arcsec$)} & \multicolumn{2}{c}{(deg)}  \\
 \multicolumn{1}{c}{(1)} & \multicolumn{1}{c}{(2)} & \multicolumn{1}{c}{(3)} & \multicolumn{1}{c}{(4)}  & \multicolumn{1}{c}{(5)} & \multicolumn{2}{c}{(6)}  & \multicolumn{2}{c}{(7)} & \multicolumn{1}{c}{(8)}  & \multicolumn{1}{c}{(9)} & \multicolumn{2}{c}{(10)} & \multicolumn{2}{c}{(11)} & \multicolumn{2}{c}{(12)} \\
  \hline
 LBABOO\,J143514.80+314052.3 & $218.81167$ & $   0.2$ & $ 31.68122$  & $   0.1$ & $18395.2$ & $ 231.8$ & $4315.1$ & $  33.6$   & $ 33.63$  & $  3$ & $  41.6$ & $   0.5$ & $  14.1$ & $   0.2$ & $   8.4$ & $   0.9$\\
 LBABOO\,J144102.55+353046.0 & $220.26061$ & $   0.0$ & $ 35.51280$  & $   0.0$ & $11964.8$ & $  56.2$ & $8396.9$ & $   7.3$   & $  7.32$  & $  7$ & $  24.4$ & $   0.0$ & $  10.1$ & $   0.0$ & $ 168.5$ & $   0.1$\\
 LBABOO\,J143410.44+331144.9 & $218.54351$ & $   0.1$ & $ 33.19582$  & $   0.3$ & $7139.8$ & $ 137.2$ & $1350.0$ & $   4.3$   & $  4.29$  & $  8$ & $  62.4$ & $   0.8$ & $  19.7$ & $   0.2$ & $ 106.9$ & $   0.9$\\
 LBABOO\,J143849.07+335015.5 & $219.70445$ & $   0.0$ & $ 33.83765$  & $   0.0$ & $6743.7$ & $  33.5$ & $4838.4$ & $   4.5$   & $  4.52$  & $  8$ & $  16.8$ & $   0.0$ & $  11.9$ & $   0.0$ & $   3.1$ & $   0.3$\\
 LBABOO\,J143318.03+345102.5 & $218.32514$ & $   0.4$ & $ 34.85072$  & $   0.4$ & $4744.6$ & $  45.5$ & $ 997.9$ & $   4.9$   & $  4.94$  & $  6$ & $  94.9$ & $   1.3$ & $  13.6$ & $   0.2$ & $ 133.8$ & $   0.8$\\
 LBABOO\,J143334.56+320908.6 & $218.39402$ & $   0.1$ & $ 32.15239$  & $   0.2$ & $3295.6$ & $  39.3$ & $ 697.0$ & $   4.5$   & $  4.48$  & $  6$ & $  41.2$ & $   0.6$ & $  22.8$ & $   0.3$ & $  70.1$ & $   1.5$\\
 LBABOO\,J142700.25+341202.0 & $216.75104$ & $   0.1$ & $ 34.20057$  & $   0.2$ & $3114.7$ & $  15.5$ & $ 410.8$ & $   1.3$   & $  1.30$  & $  9$ & $  78.9$ & $   0.5$ & $  22.8$ & $   0.1$ & $  46.8$ & $   0.4$\\
 LBABOO\,J142519.37+320707.0 & $216.33072$ & $   0.1$ & $ 32.11863$  & $   0.2$ & $2978.8$ & $ 120.7$ & $ 768.1$ & $  10.9$   & $ 10.95$  & $  2$ & $  27.6$ & $   0.5$ & $  14.9$ & $   0.3$ & $  69.9$ & $   1.9$\\
 LBABOO\,J143849.43+341553.3 & $219.70596$ & $   0.0$ & $ 34.26481$  & $   0.0$ & $2930.8$ & $  34.3$ & $1533.8$ & $   4.9$   & $  4.89$  & $  6$ & $  18.6$ & $   0.1$ & $  13.5$ & $   0.1$ & $ 143.1$ & $   0.9$\\
\vdots & \\[+0.5em]
 LBABOO\,J142926.62+334405.8 & $217.36094$ & $   2.4$ & $ 33.73496$  & $   1.1$ & $   3.0$ & $   1.3$ & $   2.9$ & $   0.7$   & $  0.75$  & $  1$ \\
 LBABOO\,J143250.18+340241.5 & $218.20909$ & $   1.8$ & $ 34.04486$  & $   1.2$ & $   3.2$ & $   1.3$ & $   3.0$ & $   0.7$   & $  0.68$  & $  1$ \\
 LBABOO\,J142755.08+334853.9 & $216.97950$ & $   2.1$ & $ 33.81498$  & $   1.4$ & $   3.4$ & $   1.5$ & $   3.1$ & $   0.8$   & $  0.77$  & $  1$ \\
 LBABOO\,J143050.88+334700.8 & $217.71200$ & $   2.1$ & $ 33.78358$  & $   1.6$ & $   3.4$ & $   1.5$ & $   3.0$ & $   0.8$   & $  0.76$  & $  1$ \\
 LBABOO\,J143128.32+335759.4 & $217.86799$ & $   1.2$ & $ 33.96652$  & $   0.6$ & $   3.9$ & $   1.0$ & $   4.5$ & $   0.6$   & $  0.66$  & $  1$ \\
 LBABOO\,J143229.21+342436.3 & $218.12169$ & $   0.9$ & $ 34.41009$  & $   0.9$ & $   3.9$ & $   1.1$ & $   4.5$ & $   0.7$   & $  0.70$  & $  1$ \\
 LBABOO\,J142950.63+342049.6 & $217.46098$ & $   2.7$ & $ 34.34712$  & $   1.3$ & $   4.1$ & $   1.6$ & $   2.9$ & $   0.7$   & $  0.71$  & $  1$ \\
 LBABOO\,J142717.32+350125.2 & $216.82216$ & $   1.9$ & $ 35.02368$  & $   1.5$ & $   4.1$ & $   1.6$ & $   3.7$ & $   0.9$   & $  0.86$  & $  1$ & $  10.3$ & $   5.0$ & $   0.0$ & $   2.9$ & $ 120.9$ & $  39.2$\\
 LBABOO\,J143541.31+334228.0 & $218.92214$ & $   1.6$ & $ 33.70780$  & $   2.0$ & $   4.1$ & $   1.7$ & $   3.3$ & $   0.9$   & $  0.83$  & $  1$ & $   9.6$ & $   4.9$ & $   3.6$ & $   3.7$ & $  18.2$ & $  78.4$\\
 LBABOO\,J143441.07+350146.7 & $218.67111$ & $   2.0$ & $ 35.02964$  & $   1.5$ & $   4.1$ & $   1.7$ & $   3.6$ & $   0.9$   & $  0.89$  & $  1$ \\
 \hline
 \end{tabular} 
 \tablefoot{\\
(1) Source name\\
(2, 3) flux-weighted position right ascension, RA, and uncertainty\\
(4, 5) flux-weighted position declination, Dec, and uncertainty\\
(6) integrated source flux density and uncertainty\\
(7) peak intensity and uncertainty\\
(8) local rms noise\\ 
(9) the number of Gaussians fitted
to the source\\
(10--12) fitted shape parameters: deconvolved
major- and minor-axes, and position angle, for extended sources, as determined by \pybdsf{}
}
\end{small}
\end{sidewaystable*}

\subsection{Astrometric precision}
We evaluated any source position errors or offsets induced by phase calibration errors by comparing the positions of sources in the LBA image with those in the  $6${\arcsec}-resolution deep HBA image of the Bo\"otes field \citep{Tasse2021}. This image has an astrometric accuracy of $0.2$\,\arcsec{} with positions corrected in the imaging process relative to the optical Pan-STARRS catalogue \citep{Chambers2016}.  We selected a sample of compact single-Gaussian sources with peak flux densities at least $7.5\sigma$ and smaller than $25$\,{\arcsec}. 

\begin{figure}
 \centering
 \includegraphics[width=0.495\textwidth]{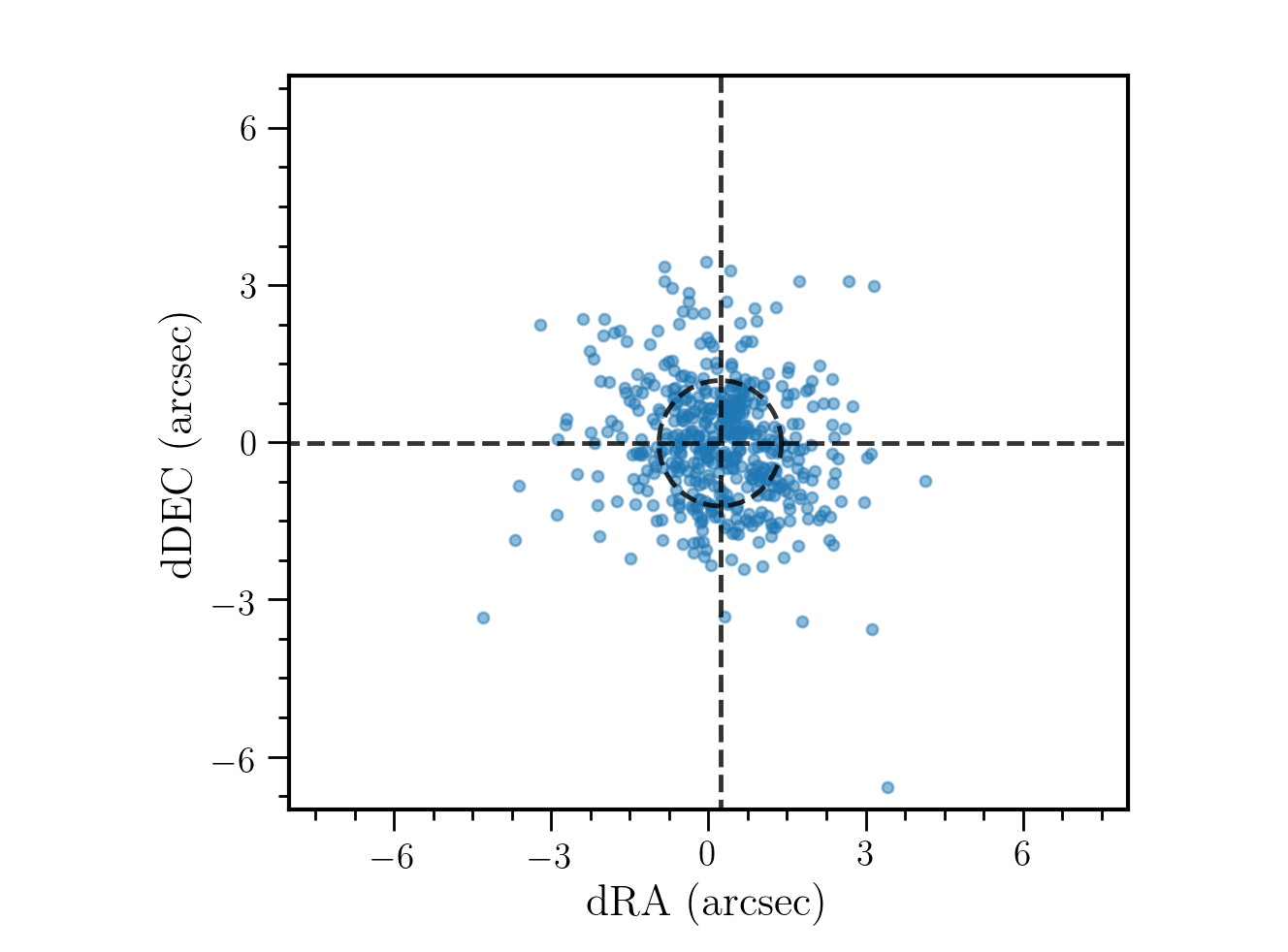}
\caption{Positional offsets between LBA and deep HBA sources. The oval shows the standard deviation of the distribution in the RA and Dec directions, where the offsets are defined as $dRA = \alpha_{\mathrm{LBA}}-\alpha_{\mathrm{HBA}}$ and $dDEC = \delta_{\mathrm{LBA}}-\delta_{\mathrm{HBA}}$, respectively.}
\label{fig:posoffset}
\end{figure}
\begin{figure}
 \centering
\includegraphics[width=0.495\textwidth]{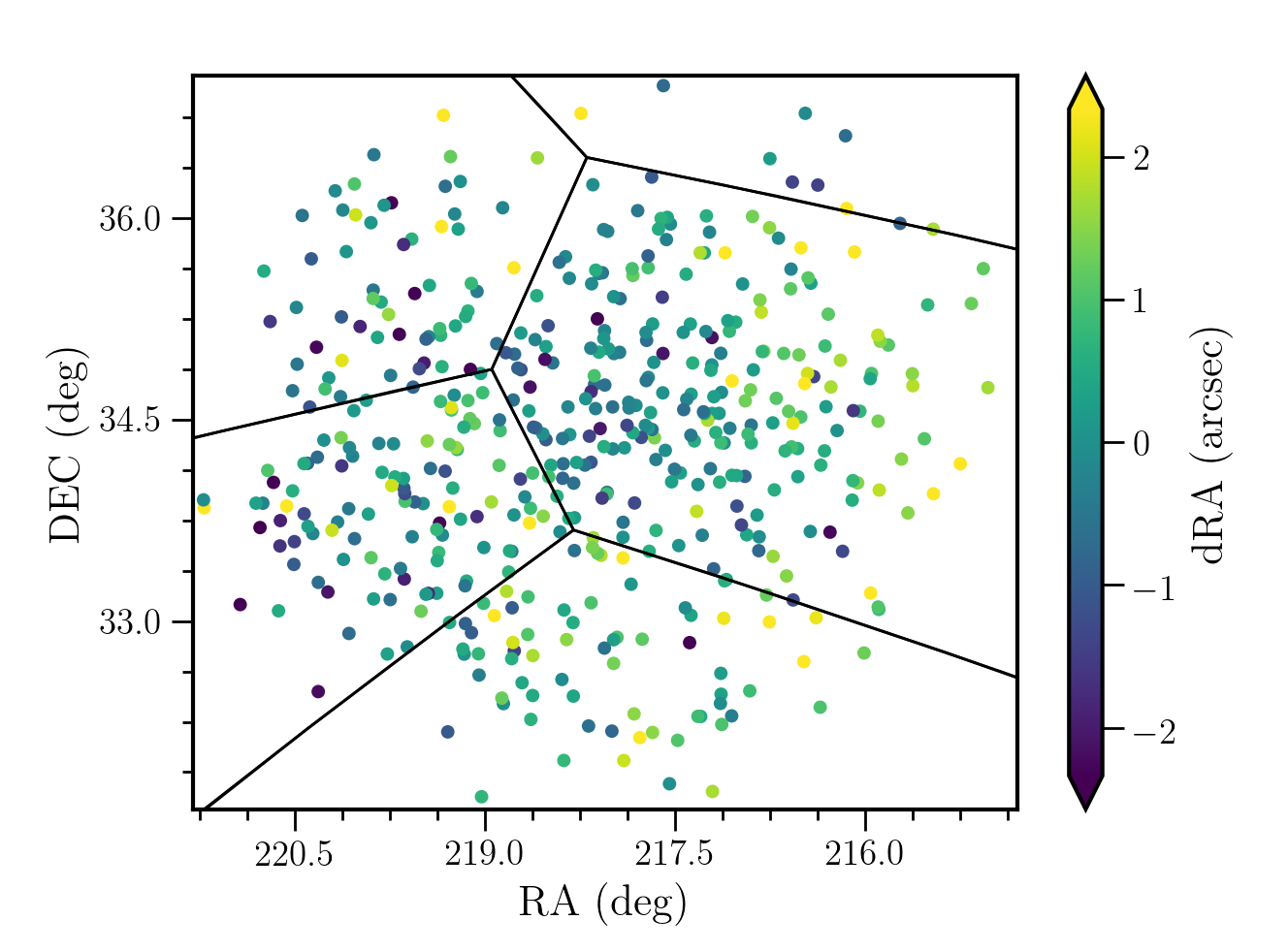}\\
 \includegraphics[width=0.495\textwidth]{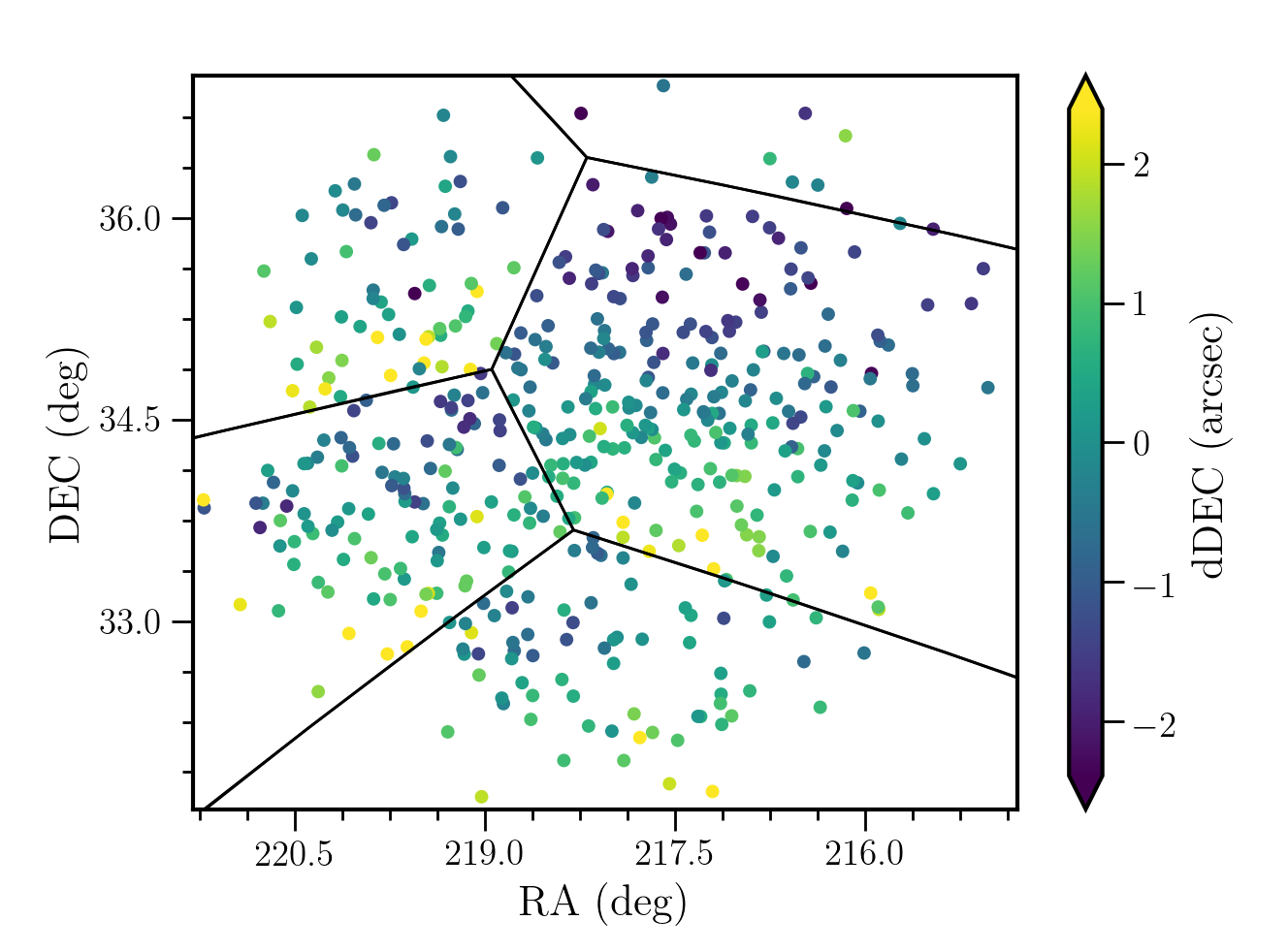}
\caption{Positional offsets between LBA and deep HBA sources in right ascension, $dRA = \alpha_{\mathrm{LBA}}-\alpha_{\mathrm{HBA}}$, (top) and declination, $dRA = \delta_{\mathrm{LBA}}-\delta_{\mathrm{HBA}}$, (bottom) as a function of position on the sky.  The black polygons show the facets used in the calibration and imaging.}
\label{fig:posoffset_2d}
\end{figure}

From this sample of $480$ sources, we measured small offsets between the positions of the sources in the LBA and the HBA of
 $dRA = \alpha_{\mathrm{LBA}}-\alpha_{\mathrm{HBA}} = 0.22 \pm 0.05$\arcsec{} ($\sigma=1.17$\arcsec{}) and
 $dDEC = \delta_{\mathrm{LBA}}-\delta_{\mathrm{HBA}} = -0.02 \pm 0.05$\arcsec{} ($\sigma=1.20$\arcsec{})
which is of the order of the pixel
size of the LOFAR observations and the HBA accuracy ($\sim0.2$\arcsec). The offsets are plotted in Fig.\,\ref{fig:posoffset} and in  Fig.\,\ref{fig:posoffset_2d} as a function of position on the sky. While a per-facet positional correction was made in the imaging (see Sect.\,\ref{sect:ddcal}),  a trend in the declination offsets is still apparent within each facet, in that at higher declination the LBA positions are shifted slightly south ($\sim1$\,{\arcsec}), while at lower declination, the LBA positions are shifted slightly north ($\sim1$\,{\arcsec}). This is likely due to refraction and systematic ionospheric effects. Refraction causes lines of sight at greater zenith angles through the thicker projected ionosphere to be subject to greater deflection angles. As the observations are centred on transit, the right ascension offset should average out leaving only a bulk declination offset. Deflection angle differences of a few arcseconds over a degree are possible assuming a typical ionosphere height and thickness. Furthermore, the ionosphere typically has a strong north--south gradient in electron density during the daytime that may compound this effect. Smaller facet sizes would allow these effects to be corrected during self-calibration.

\subsection{Smearing}
\label{sect:smearing}
\begin{figure}
 \centering
\includegraphics[width=0.45\textwidth]{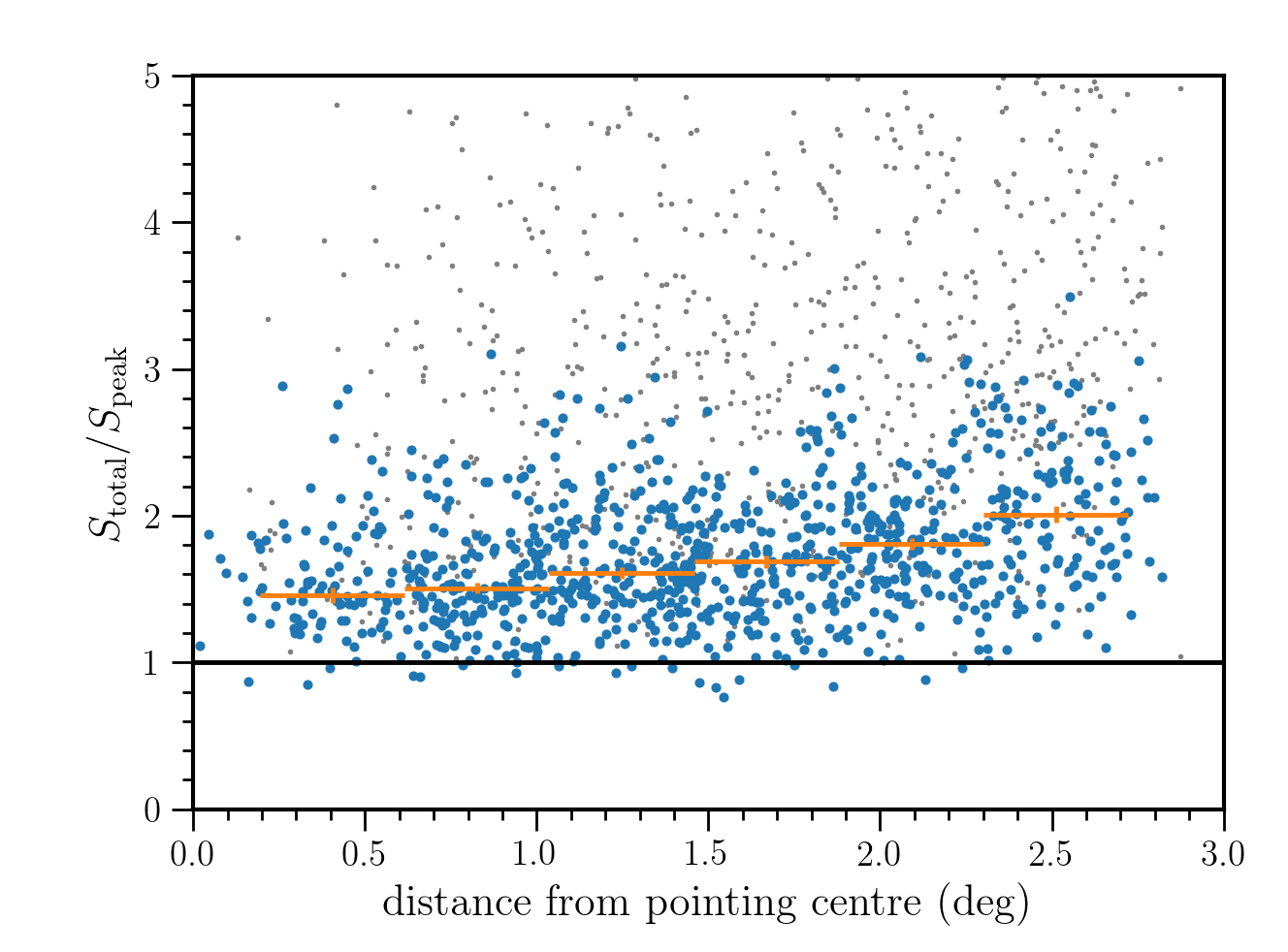}
\caption{
Measured ratio of integrated to peak flux density as a function of distance from the pointing centre {for all sources (in gray) and for a selection of compact sources (in blue)}. The orange crosses show the median smearing of compact sources, which is systematically greater than unity and increases with radius, likely because of imperfect calibration.
}
\label{fig:total_peak}
\end{figure}

 By plotting the ratio of the total flux density to the peak intensity  (see Fig.\,\ref{fig:total_peak}), it is apparent that there is some level of smearing in that this ratio exceeds unity for almost all sources. For unresolved sources, this ratio should be scattered around unity. The median ratio for small ($<30$\,\arcsec{}) single-Gaussian sources increases as a function of distance from the pointing centre, from around $1.5$ at the centre to $\sim2$ at a radius of $2.5$ degrees. As \ddfacet{} is able to account for a varying psf due to bandwidth and time-averaging during deconvolution, and all these sources are deconvolved, this is likely a result of imperfect phase calibration applied over the very large facets (see Fig.\,\ref{fig:image}). As the facets are large and spread from the centre to the edges of the field, the phase solutions are most applicable in the direction of  the greatest apparent flux density of sources which is towards the centre of the field.

\subsection{Completeness}
\label{sect:completeness}

To assess completeness, we performed a Monte-Carlo simulation in the image plane in which we generated {$25$ random fields each containing approximately $7000$ randomly positioned point sources with flux densities between $2$ and $200$\,\mJy.} The source flux densities were drawn randomly from {the $150$-MHz source count distribution of \cite{Mandal2021}, scaled to $54$\,MHz assuming an average spectral index of $-0.8$}. To simulate the effect of smearing, we increased both the major- and minor-axes of these injected 2D Gaussians by $\sqrt{S_f}$, where $S_f$ is the median smearing factor as a function of distance from pointing centre (see Fig.\,\ref{fig:total_peak}).  Sources were  added to the residual image produced by the original source detection with {\scshape PyBDSF}. For each simulated image, sources were detected with the same {\scshape PyBDSF} parameters as those described in Section~\ref{sect:cat}, but using the original rms image because this more accurately captures the increased noise level around real bright sources in the image by decreasing the box size used for determining the rms. {In each field, approximately $1500$ of the simulated sources were detected.} 

\begin{figure}
 \centering
 \includegraphics[width=0.45\textwidth]{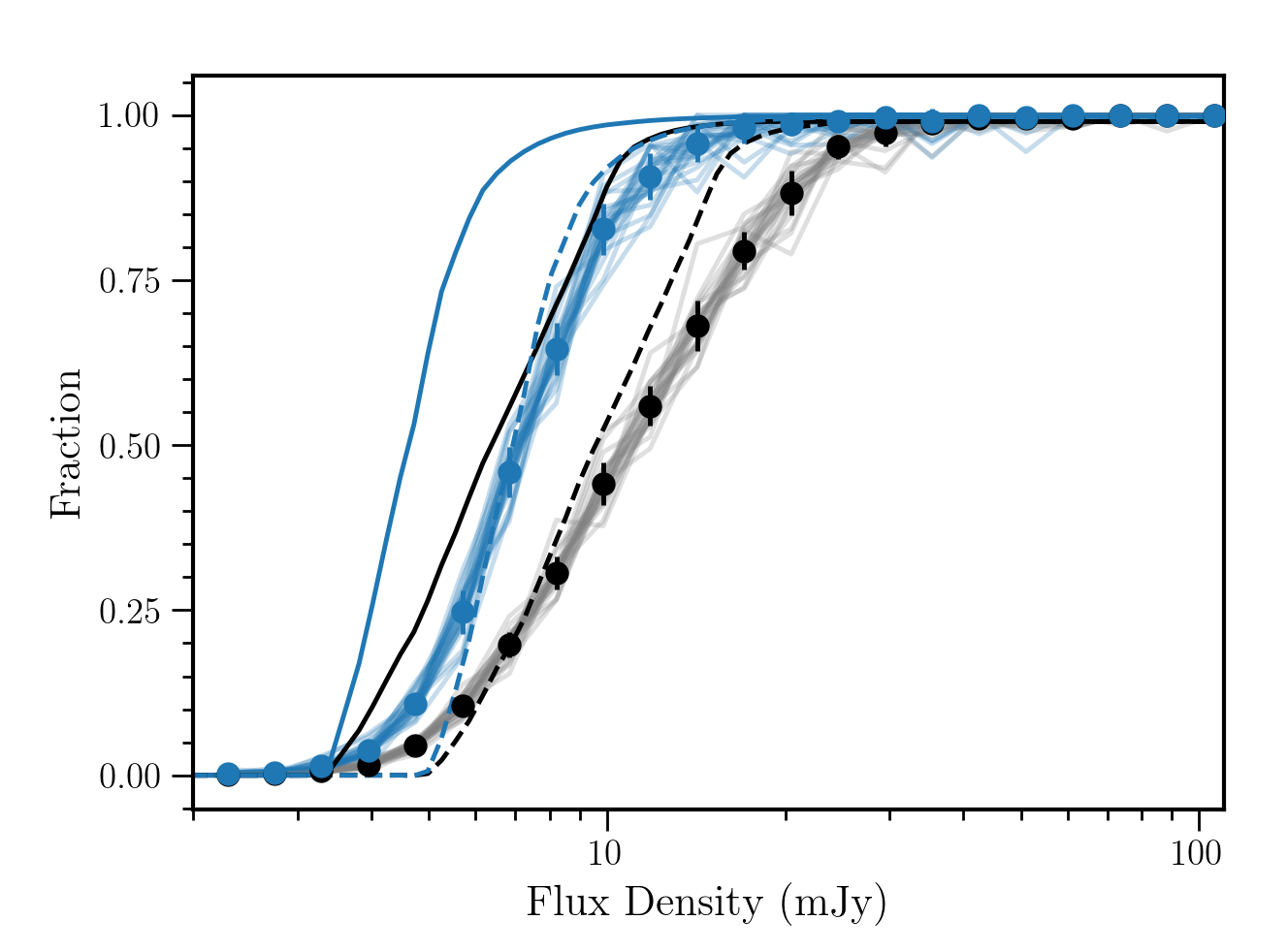}\\
 \includegraphics[width=0.45\textwidth]{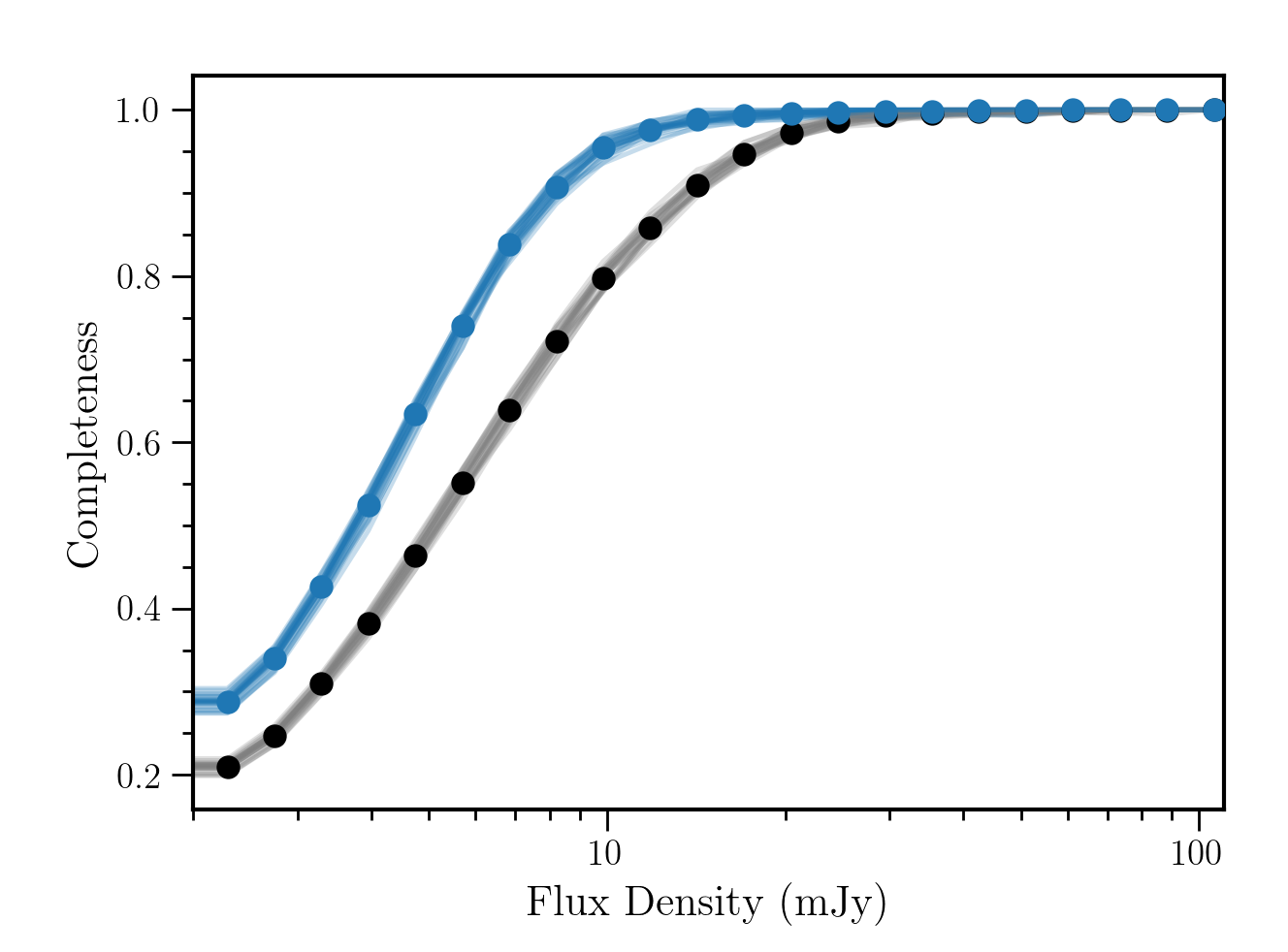}
\caption{{Top: Fraction of simulated sources detected as a function of flux density for each simulated image (in grey) and the median binned points (in black). The solid black line shows the fraction of sources that could be detected with a $5\sigma$ peak intensity threshold based on the rms only, while the dashed black line shows the same but assuming a single average smearing factor of $1.5$, most applicable for the center of the field. The blue points and lines show the same, but determined only over the area with deep optical coverage. Bottom: Estimated completeness above a given flux density ---assuming the 150-MHz source count distribution of \cite{Mandal2021}--- scaled to $54$\,MHz assuming an average spectral index of $-0.8$. This is estimated for the full field (shown in grey and black), and for the region of deep optical coverage (shown in light and solid blue).}}
\label{fig:detF}
\end{figure}

For each simulated image, we determined the fraction of sources detected as a function of flux density (shown in Fig.\,\ref{fig:detF}). Sources are considered detected by matching their coordinates with the injected source catalogue. Only about $250$ sources in each field satisfy the detection criterion of peak intensity $>5\sigma$.  Due to the smearing, the detected fraction in the simulations deviates from  the fraction of point sources that would be detected based purely on the fraction of the rms below a given value. The completeness, that is,  the fraction of recovered sources above a given flux density, is also shown in Fig.\,\ref{fig:detF}. This indicates that the catalogue covering the full field is around 90\%\ complete at a flux density of $10$\,mJy, while that covering the deep optical area is around 90\%\ complete at a flux density of $6.5$\,mJy.

\subsection{Flux density scale}
Given the uncertainties in the low-frequency flux density scale \citep[e.g.][hereafter SH12]{ScaifeHeald2012} and the LOFAR station beam models, we may expect some systematic errors in the measured LOFAR flux densities. In this section we evaluate the uncertainties in the measured LOFAR flux densities and make corrections to the catalogue to account for systematic effects and ensure that the flux densities are on the \citetalias{ScaifeHeald2012} flux density scale.

\begin{figure}
 \centering
 \includegraphics[width=0.5\textwidth]{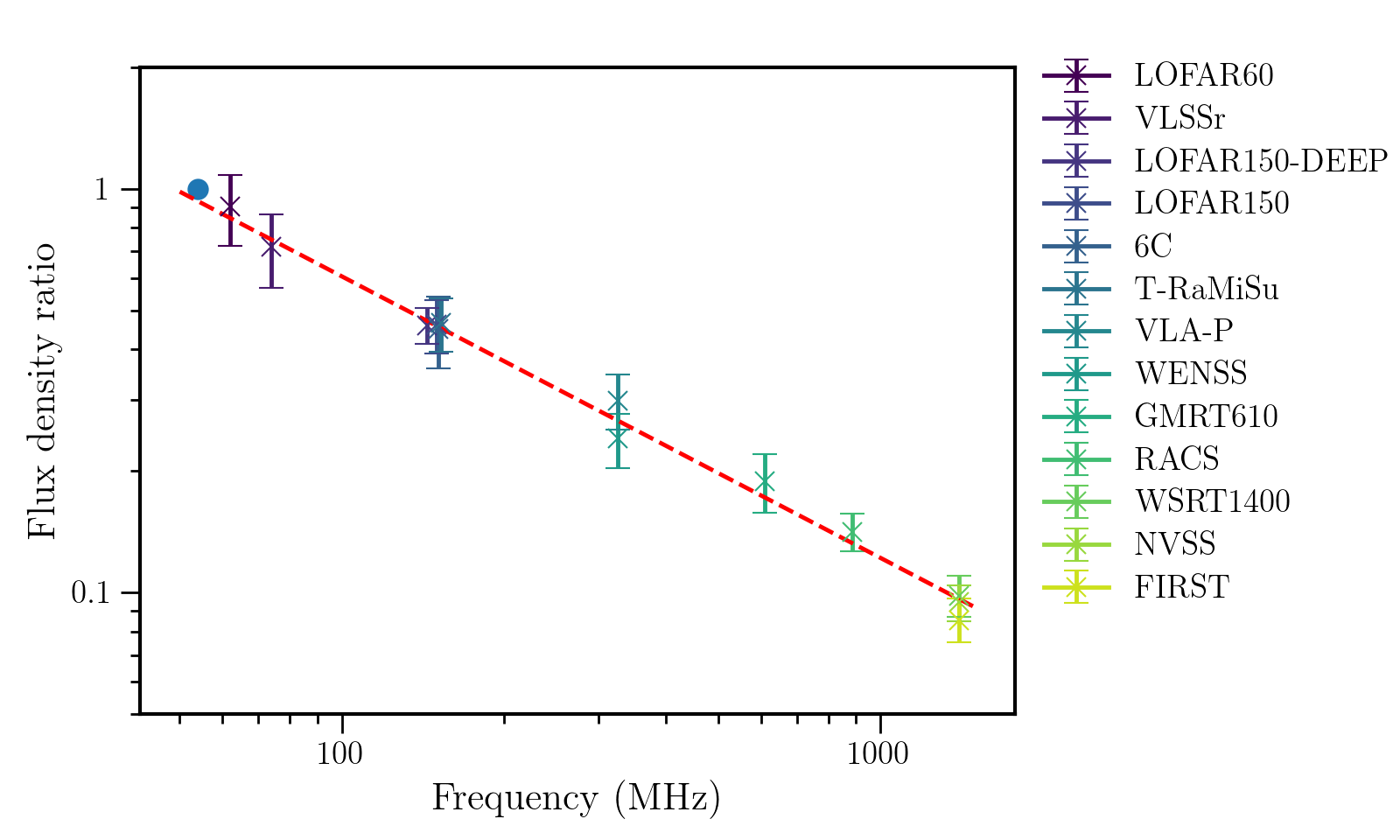}
\caption{{Flux density scaling predicted from observations at multiple frequencies compared to that observed at $54$\,MHz here.}}
\label{fig:fluxpred}
\end{figure}

We followed the method of \cite{Sabater2021} to investigate any flux density scale offset, and cross-matched the LBA catalogue with multiple catalogues at higher frequencies, including only relatively compact and isolated sources for the comparison. {The flux density measurements from the literature have been adjusted to bring them on to the \citetalias{ScaifeHeald2012} scale adopted here, which was designed to be more accurate than previous scales at frequencies below $\sim300$\,MHz. The  \citetalias{ScaifeHeald2012} scale was calibrated using data on the \citet{RCB1973} flux density scale, which is consistent with  the \citet{KPW1969} scale above $300$\,MHz. Data from the 1.4-GHz VLA surveys ---NVSS and FIRST--- are on the \citet{Baars1977} flux density scale, and so we used their correction factor \citep[cf. table 7 of ][]{Baars1977} to align the flux densities of these catalogues with the  \citet{KPW1969} scale. The 1.4-GHz WSRT data were corrected  based on NVSS and so we applied the same correction factor. The RACS data are on the \cite{Reynolds1994} flux density scale. This is consistent with the \citet{Baars1977} flux density scale and therefore we used their correction factors (interpolated to 888\,MHz from their Table 7) to align the RACS flux densities with the  \citet{KPW1969} scale. The $325$-MHz data are somewhat more complicated: the correction factor to scale data from WENSS to the \citetalias{ScaifeHeald2012} flux scale is an average correction across the discrete set of WENSS calibrators (3C 48, 3C 147, 3C 286 and 3C 295), but this should not apply to localised regions of the sky. As the Bo\"otes field is only 18\,deg from 3C\,295, it is likely that this is the calibrator that was used for this part of the sky. We therefore scaled the WENSS flux densities by the offset for 3C\,295. The 325-MHz VLA data were corrected to the WENSS flux density scale, but we further applied the additional offset factor of 0.95 found by \cite{Coppejans2015}. The deep 150-MHz data were corrected by the factor determined by  \cite{Sabater2021}. The flux densities of the remaining  catalogues were all calibrated with respect to \citetalias{ScaifeHeald2012} or \citet{RCB1973}, and so no further corrections were made.}  {Table~\ref{tab:cross} lists all the catalogues used with their flux scale corrections and the details of the cross-matching selection to select only compact isolated sources}. For the catalogues that are much deeper, we considered only the brighter sources to avoid any bias from a changing spectral index at lower flux densities (see Sect.\,\ref{sect:specind}). {Following the method of \cite{Sabater2021} we set flux density thresholds for both of the cross-matched surveys to avoid a
bias towards sources with high absolute values of their spectral indices}.  For each catalogue, we calculated the ratio of the catalogued flux densities  to those at $54$\,MHz. 
The median of this ratio is plotted in Fig.\,\ref{fig:fluxpred}. {The straight line fitted to these ratios gives a value at $54$\,MHz of  $0.96 \pm  0.11$}, suggesting that the LBA catalogued flux densities are consistent with unity. We therefore apply no correction to the catalogued flux densities.  


\begin{table*}
\caption{Catalogues and cross-match parameters used to {test} the flux density scale}
\label{tab:cross}
\begin{small}
\begin{center}
 \begin{tabular}{llllp{1cm}p{1cm}p{1.2cm}p{1.2cm}ll}
 \hline
     Catalogue   &    Frequency &  Size Limit& Resolution& Match radius& Flux Limit& {Flux scale correction\tablefootmark{a}}& Flux scale error & Flux Ratio & Reference \\
        &    (MHz)&  (arcsec) & (arcsec)& (arcsec)& (mJy)&  &  &  & \\
\hline \\[-0.9em]
      LOFAR60 &  62 &     60 &      30 &   30 &      1 & 1  &         0.2 & $0.962_{0.193}^{0.195} $  & 1 \\[+0.2em]
        VLSSr &  74 &     30 &      75 &   75 &    530 &  1 &         0.2  & $0.787_{0.159}^{0.170}  $ & 2 \\[+0.2em] 
           6C & 151 &     30 &      60 &   60 &    250 & 1  &         0.2  & $0.517_{0.106}^{0.105}  $ & 3 \\[+0.2em]
     LOFAR150 & 150 &     60 &       7.5 &   15 &     0.01 & 1  &        0.15  & $0.463_{0.070}^{0.070} $ & 4 \\[+0.2em]
     T-RaMiSu & 153 &     30 &      25 &   15 &     10 & 1  &        0.15  & $0.503_{0.078}^{0.076}  $ & 5 \\[+0.2em]
LOFAR150-DEEP & 144 &     60 &       6 &    6 &     10 & 0.859  &         0.1  & $0.450_{0.045}^{0.046}  $  & 6 \\[+0.2em]
        VLA-P & 325 &     30 &       5.5 &   15 &      2 & 0.935  &        0.15  & $0.319_{0.048}^{0.050} $  & 7 \\[+0.2em]
        WENSS & 325 &     60 &      30 &   30 &     15 & 0.982   &       0.15  & $0.250_{0.038}^{0.038}   $ & 8 \\[+0.2em]
      GMRT610 &  610 &     30 &       5.5 &    5.5 &      2 & 1  &        0.15  & $0.207_{0.038}^{0.036} $ & 9  \\[+0.2em]
      RACS &  888 &     30 &       25 &    15 &      2 & 1  &        0.15  & $0.207_{0.038}^{0.036} $ & 10  \\[+0.2em]
     WSRT1400 & 1400 &     60 &      54 &   54 &      1 & 1.029 &         0.1  & $0.100_{0.012}^{0.013}  $  & 11 \\[+0.2em]
         NVSS & 1400 &     60 &      45 &   45 &      2 & 1.029  &        0.1  & $0.092_{0.010}^{0.010}  $ & 12 \\[+0.2em]
        FIRST&  1400 &     30 &       5 &   15 &      2 & 1.029  &        0.1  & $0.091_{0.010}^{0.010}  $ & 13 \\[+0.2em]
 \hline
 \end{tabular}
 \tablefoot{\\
 \tablefootmark{a} value by which the flux densities in the catalogue were multiplied to bring them onto the \citetalias{ScaifeHeald2012} flux density scale -- see details in the text.}
\tablebib{
(1)~\citet{vanWeeren2014}; (2) \citet{Lane2014}; (3) \citet{Hales1988}; (4) \citet{Williams2016};
(5) \citet{Williams2013}; (6) \citet{Tasse2021}; (7) \citet{Coppejans2015}; (8) \citet{Rengelink1997}; (9)  \citet{Coppejans2016}; (10) \cite{McConnell2020}; (11) \citet{deVries2002}; (12) \citet{Condon1997}; (13) \citet{Becker1995}.
}
\end{center}
\end{small}
\end{table*}

\begin{figure}
 \centering
 \includegraphics[width=0.495\textwidth]{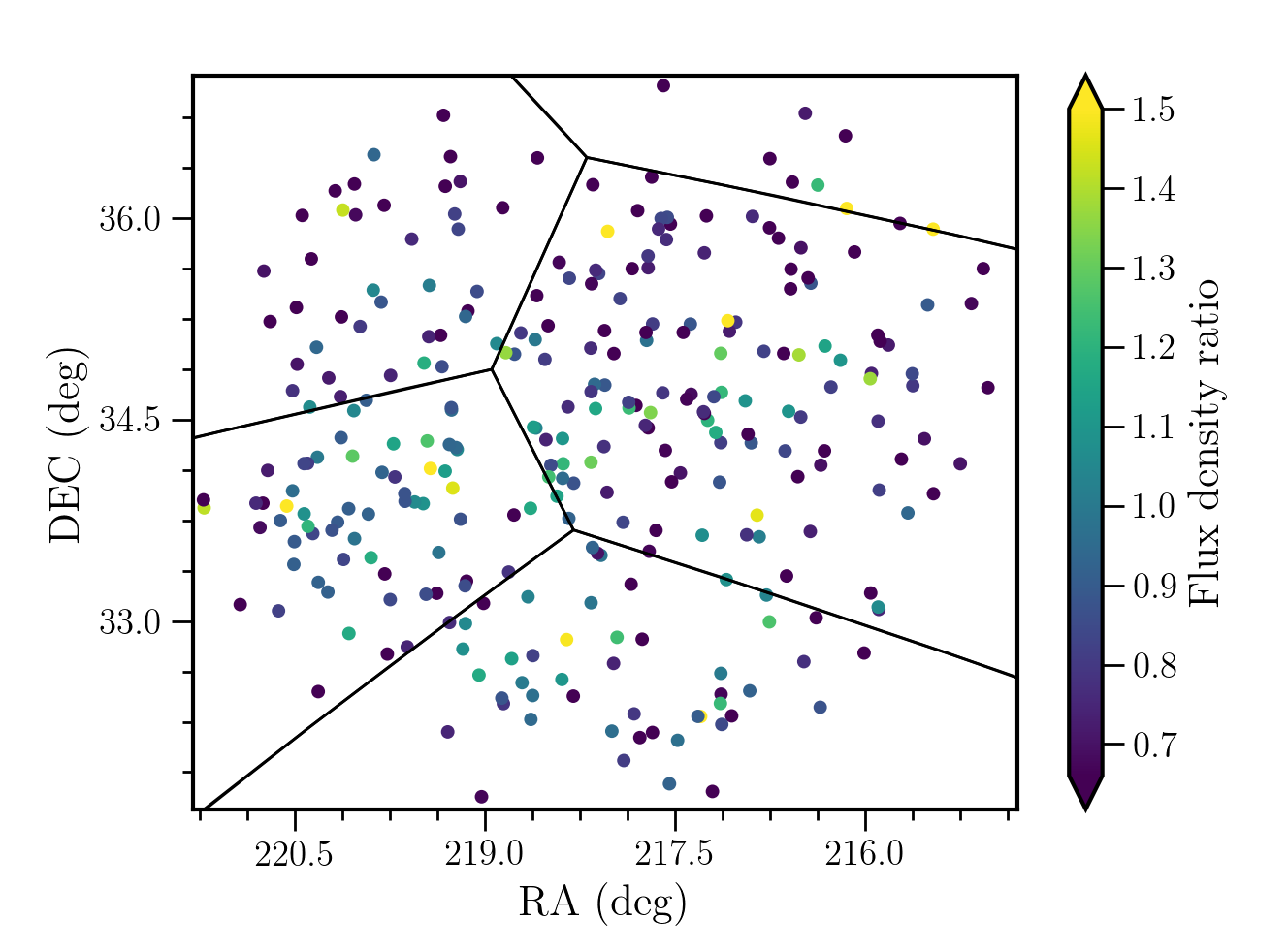}
\caption{Ratio of flux density at $54$\,MHz and the scaled $144$-MHz flux density as a function of position on the sky. The black polygons show the facets used in the calibration and imaging.}
\label{fig:fluxratiopos}
\end{figure}

We also checked for any variation in the total flux density ratio with distance from the phase centre or position on the sky and found none. The ratio between the $54$-MHz and $144$-MHz flux densities is shown as a function of sky position in Fig.~\ref{fig:fluxratiopos} for a sample of bright ($S_{54 \mathrm{MHz}}>20$\,mJy), high-signal-to-noise-ratio ($S/N>7.5$), and small ($<30$\,\arcsec{}) sources. Here, the $144$-MHz flux densities from the deep HBA image have been scaled to $54$\,MHz with an assumed spectral index of $-0.7$. There is no noticeable trend with position.

\section{Spectral indices and source counts}

\label{sect:results}
The $1,948$ sources in the catalogue provide a statistically
significant sample across three orders of magnitude in flux
density from $3$\,mJy to $18$\,Jy. The addition of the deep $144$-MHz data available allows the calculation of spectral indices for all sources in the catalogue. In this section we present the derived spectral index distributions and 54-MHz source counts.

\subsection{HBA cross-matching}
\label{sect:deepmatch}
Through a process of visual inspection we matched the LBA sources with those from the deep $144$-MHz HBA image \citep{Tasse2021}, which has a central noise level of  $30$\,\muJybeam{}. Within the area of deep optical coverage, we performed the cross-match against the catalogue of \cite{Kondapally2021}, which includes the optical identifications as well as the restructuring of the raw \pybdsf{} source catalogue from \citet{Tasse2021} into true physical sources. In this process, following that of \cite{Williams2016} and \cite{Kondapally2021}, we decomposed some of the LBA \pybdsf{} sources into their Gaussian components and (re-)combined some \pybdsf{} (Gaussian components) sources into final physical sources. We also flagged some \pybdsf{} sources as artefacts. The resulting matched catalogue contains $1,789$ sources ---of which $829$ ($46$\,per\,cent) lie within the deep optical coverage--- and is available online\textsuperscript{\ref{note1}}. The much greater depth of the HBA data  (about 50 times greater for a source of spectral index $-0.7$) means that all the LBA sources are detected in the HBA image; a source with a $54$-MHz flux density of $5$\,mJy would have to have a spectral index steeper than $-2.5$ to be undetected at $144$\,MHz and there are no such sources in the field.

\subsection{Spectral Indices}
\label{sect:specind}

\begin{figure}
 \centering
 \includegraphics[width=0.495\textwidth]{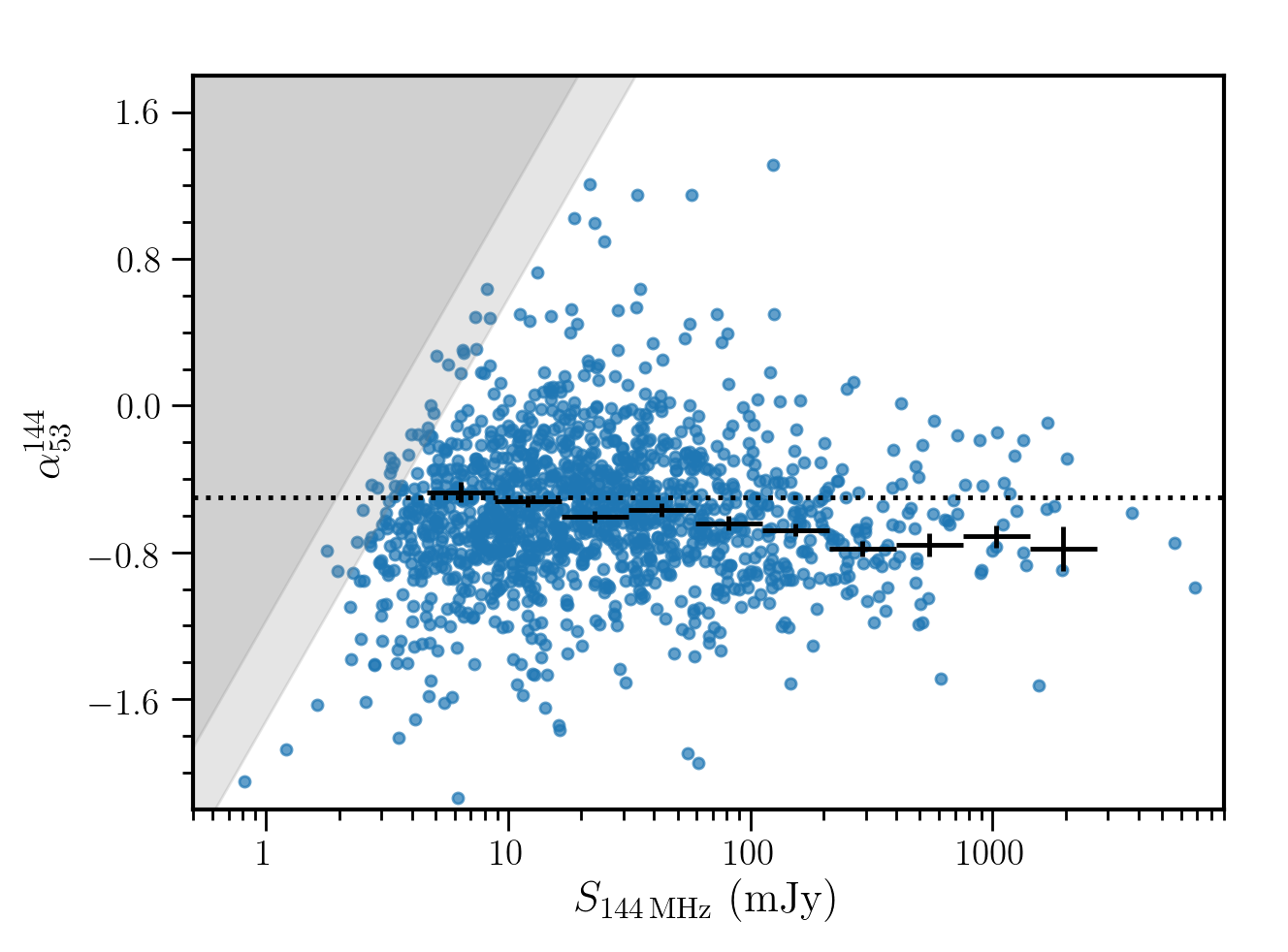}
\caption{Spectral indices measured between $54$ and $144$\,MHz with LOFAR LBA and HBA respectively. The grey shaded areas show where the distribution becomes incomplete for flatter sources due to the varying sensitivity of the LBA observations, assuming a $5\sigma$ detection of a point source at the minimum ($0.65$\,\mJybeam{}, light grey) and median ($1.1$\,\mJybeam{}, dark grey) rms. The dotted line shows a value of $-0.5$ to guide the eye. The black points show the median values and errors on the median within equally spaced logarithmic flux-density bins, showing a flattening of spectra towards lower flux densities.}
\label{fig:specind}
\end{figure}

 The spectral indices, $\alpha_{54}^{144}$, calculated for these matched sources are plotted in  Fig.\,\ref{fig:specind} as a function of their HBA flux densities.  The greater depth of the HBA data means that the faintest HBA sources also detected at $54$\,MHz preferentially have steeper spectra, that is, the faintest HBA sources with flatter spectra fall below the detection limit of the LBA. This incompleteness is indicated by the shaded grey areas in Fig.\,\ref{fig:specind} where sources within the darkest region cannot be detected and those within the lighter region may only be detected in part of the LBA image. We therefore only calculate the median spectral indices in bins above 144-MHz flux densities of $5$\,\mJybeam{}. Nevertheless, there is a clear trend towards flatter spectral indices with decreasing $144$-MHz flux density, from $\sim-0.75$ at flux densities between $100$ and $1000$\,mJy to $\sim-0.5$ at flux densities between $5$ and $10$\,mJy.

\begin{figure}
 \centering
 \includegraphics[width=0.495\textwidth]{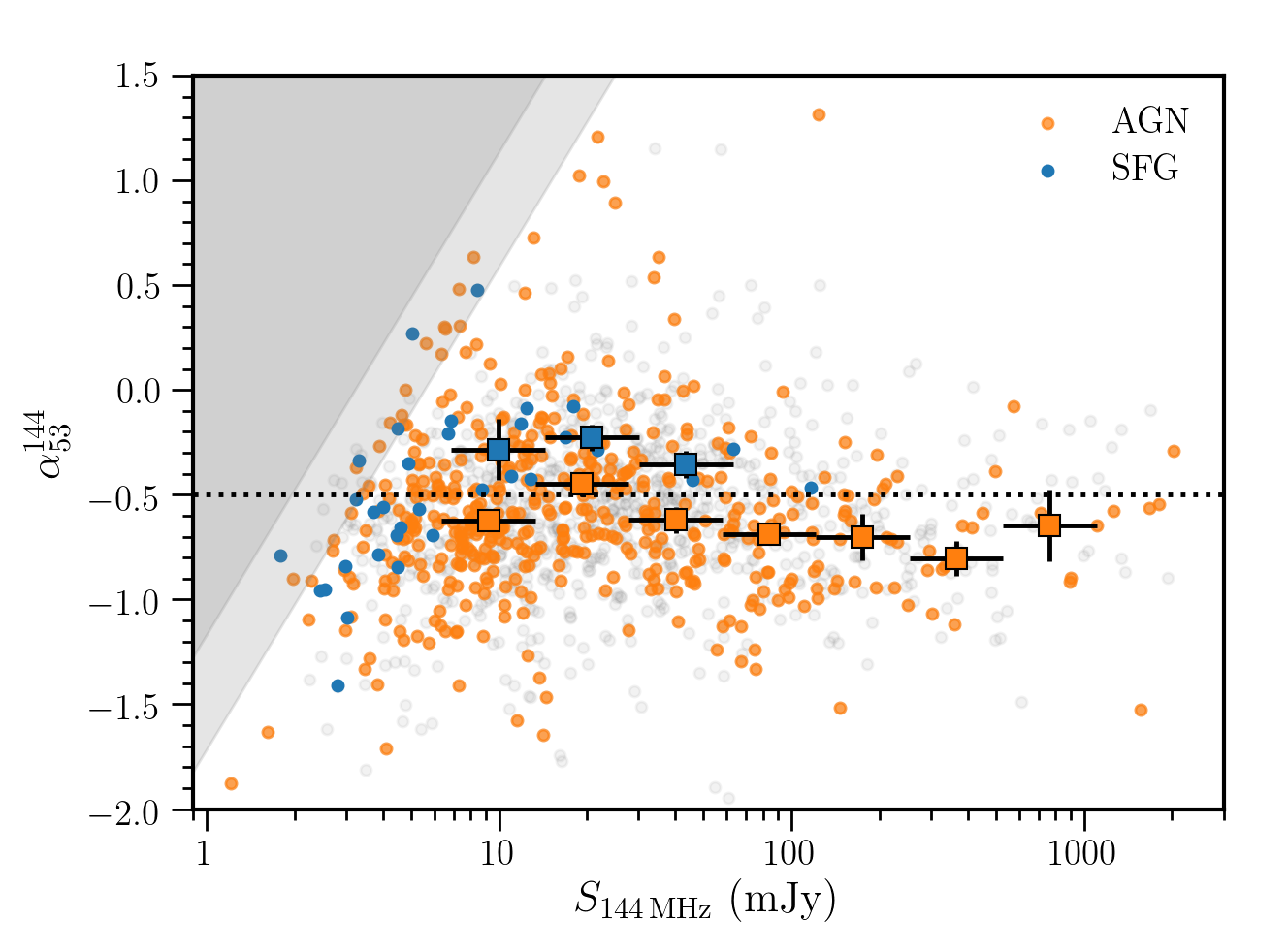}
\caption{Spectral indices measured between $54$ and $144$\,MHz, with LOFAR LBA and HBA respectively. Grey points show all matched sources, while the coloured points show those within the optical coverage, where sources have been classified as SFGs (blue) or AGN (orange). The large coloured points show the median values and errors on the median within equally spaced logarithmic flux density bins for the two populations separately.}
\label{fig:specindtype}
\end{figure}

Using the optical identifications \citep{Kondapally2021} and source classifications (Best et al. in prep) available for the deep HBA sources, we investigated the spectral indices for AGN and SFGs separately. Of the $829$ LBA sources within the deep optical coverage, $61$ are classified as SFGs, while $626$ are classified as radio-loud AGN, and $29$ as radio-quiet AGN. A further $113$ are unclassified. These fractions are broadly consistent with predictions from the SKA simulated skies models \citep[SKADS][]{Wilman2008}, which yield about $30$ SFGs and about $800$ AGN taking into account the varying rms and masked deep optical area. The differences may be a result of different models used within SKADS or the SFG/AGN classification used by Best et al. (in prep) which relies on the radio excess above the far-infrared radio correlation.   The spectral indices for the SFGs and radio-loud AGN are shown in Fig.\,\ref{fig:specindtype}, and again only showing the median spectral indices in $144$-MHz flux density above $5$\,\mJybeam{}, where the median values are not affected by the incompleteness.  The majority of the sources are classified as AGN (both high excitation and low excitation radio galaxies) with SFGs only appearing at lower flux densities, but it is apparent that the SFGs have flatter radio spectra. The median spectral index for the SFGs is $-0.4$. This is consistent with a further flattening of the spectra of SFGs compared to that found by \cite{CalistroRivera2017} at higher frequencies where the average spectral index of SFGs between $150$ and $325$\,MHz was found to be $-0.6$ while that between $325$ and $1400$\,MHz was a steeper $-0.78$ due to free-free absorption at the lower frequencies \citep[see also][]{Ramasawmy2021}.

\begin{figure}
 \centering
 \includegraphics[width=0.495\textwidth]{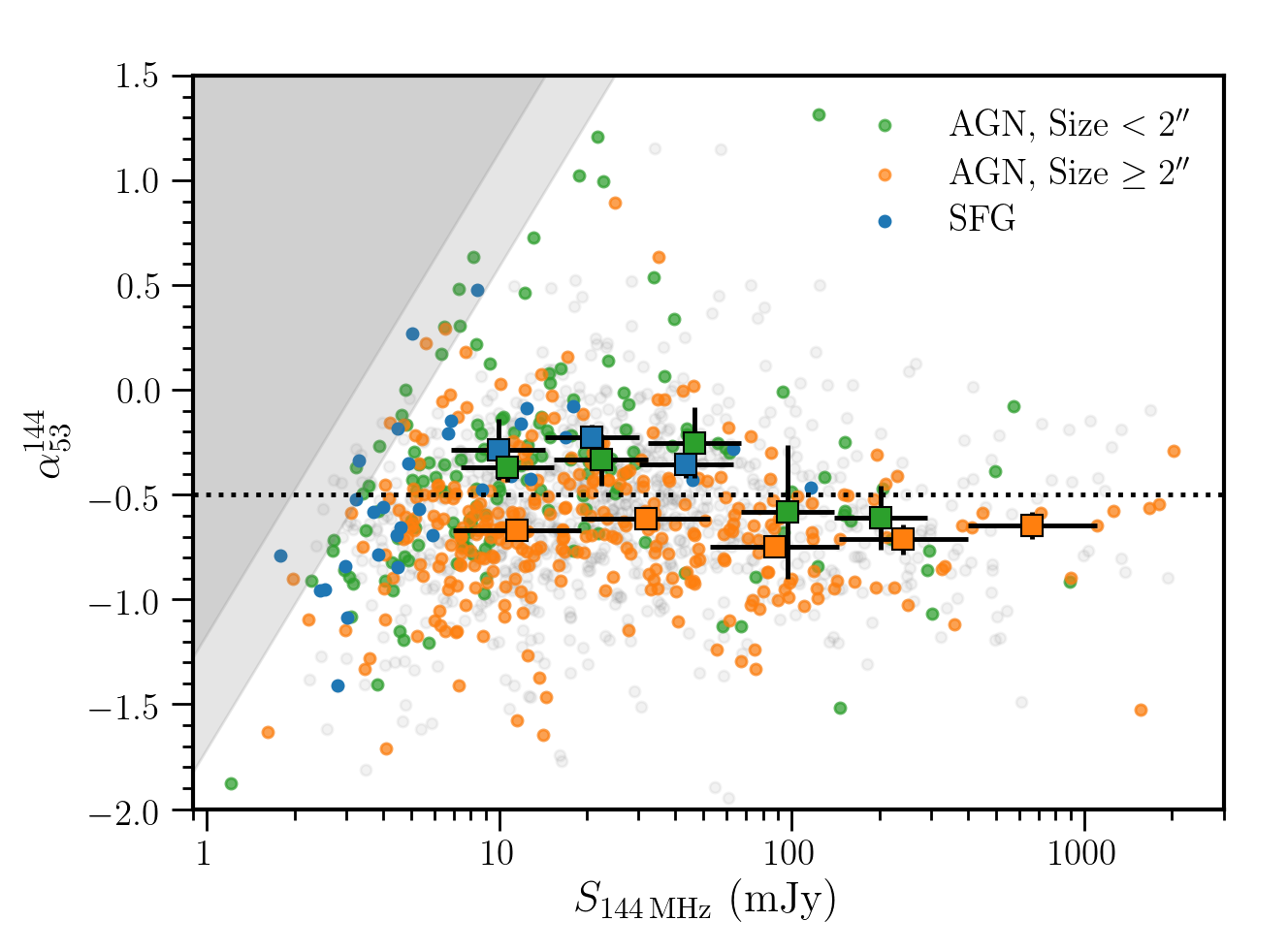}
\caption{As in Fig.\ \ref{fig:specindtype}, but only the AGN are plotted and are plotted separately for extended (orange) and compact (green) sources, where the source size is determined from the 144\,MHz data.}
\label{fig:specindtype2}
\end{figure}

The flattening of the spectral index towards lower flux densities is also evident in the AGN population. With the size information from the higher resolution HBA data, we consider the spectral indices of the AGN for about $150$ compact sources, defined as having a deconvolved size of $<2$\,\arcsec{}, and for about $320$ extended sources of $\geq2$\,\arcsec{}. This is shown in  Fig.\,\ref{fig:specindtype2}, from which it is apparent that the compact sources have flatter median spectral indices at lower flux densities compared to the more extended sources, which remain steep ($\alpha^{144}_{54}\sim-0.7$). This is similar to the result of \cite{deGasperin2018a} who found a similar trend in the spectral index between $150$\,MHz and $1.4$\,GHz with both flux density and source compactness. This is likely because the population of large sources is dominated by lobe-dominated sources, where more synchrotron emission comes from old electron populations and is thus steeper. The overall flattening of the spectral indices at lower flux densities is therefore likely driven by a growing population of SFGs and core-dominated, compact AGN.

\subsection{Source counts}
\label{sect:srccnt}
\begin{figure}
 \centering
\includegraphics[width=0.495\textwidth]{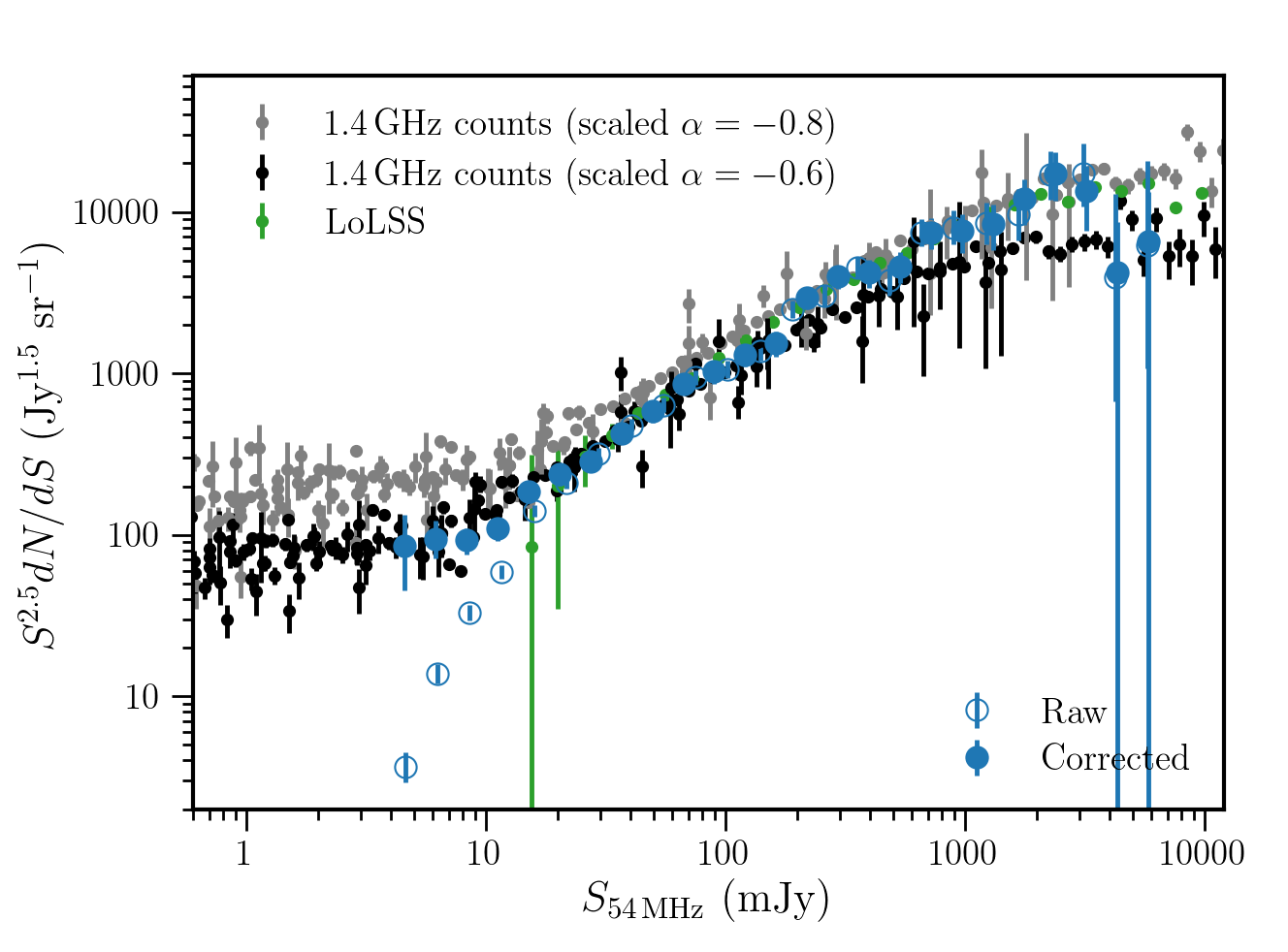}
\caption{{Euclidean-normalised differential source counts for
the Bo\"otes field between 4\,mJy and 7\,Jy}. The open circles show the
raw, uncorrected source counts, while the filled circles show the completeness-corrected
counts. For comparison we include the
$1.4$\,GHz source counts from various surveys compiled by \citet{deZotti2010}, \citet{Bonato2017}, and \citet{Bonato2021}, and scaled to $54$\,MHz assuming a spectral index of
$-0.8$ (in gray) and $-0.6$ (in black), and the $54$-MHz LoLSS source counts between 10\,mJy and 30\,Jy (small green points).}
\label{fig:srccnts}
\end{figure}

The Euclidean-normalised differential
source counts for this catalogue are plotted in Fig.\,\ref{fig:srccnts}, where we show both the raw counts and those corrected for completeness using the results of Fig.\,\ref{fig:detF}.  The errors on the raw counts per flux density bin are the Poisson errors corrected for small numbers \citep{Gehrels1986}. The primary causes of incompleteness are the varying rms level due to the `primary' beam (see Fig.\,\ref{fig:mosrms}) and the residual ionospheric smearing (see Sect.\,\ref{sect:smearing}). 
{To account for this, we used the detection fraction (see Fig.\,\ref{fig:detF}) determined from the completeness simulations to correct the raw source counts. Errors on the final counts are propagated from the errors on the detection fraction from the simulations. The source counts presented here were determined using the full image, but agree with those determined only within the area of deep optical coverage or within $1$\,degree of the phase centre, where the smearing is less dominant and the completeness is higher. It should be noted that the lowest flux-density bins are determined from sources only found in the central region ($\lesssim3$\,\sqdeg{}), and so they may be affected by cosmic variance. }
Following the method outlined by \cite{Prandoni2001} and \cite{Williams2016}, we make a correction for the resolution bias, that is, the preferential non-detection of large sources, which takes into account the size distribution of sources. This correction is much smaller than the completeness correction. The counts presented here complement the LoLSS source counts \citep{deGasperin2021}, which provide better statistics at the brighter end, while our deep-field counts probe the fainter sources where LoLSS becomes incomplete.

We compared the Euclidean-normalised source counts derived here with the  compilations of $1.4$\,GHz source counts  by \citet{deZotti2010}, \citet{Bonato2017}, and \citet{Bonato2021}. We find very good agreement with the higher-frequency counts if we assume different spectral indices at different flux densities. There is some deviation in the lowest flux density bin $4$--$5$\,mJy, which may be a result of incompleteness. An average spectral index from $-0.8$ works well at $54$-MHz flux densities above $\sim100$\,mJy, while an average spectral index of $-0.6$ works well at lower flux densities. This is consistent with the change in the measured median spectral indices of individual sources both in this work (see Sect.\,\ref{sect:specind}) and that of  \cite{deGasperin2018a}.

\begin{figure}
 \centering
\includegraphics[width=0.495\textwidth]{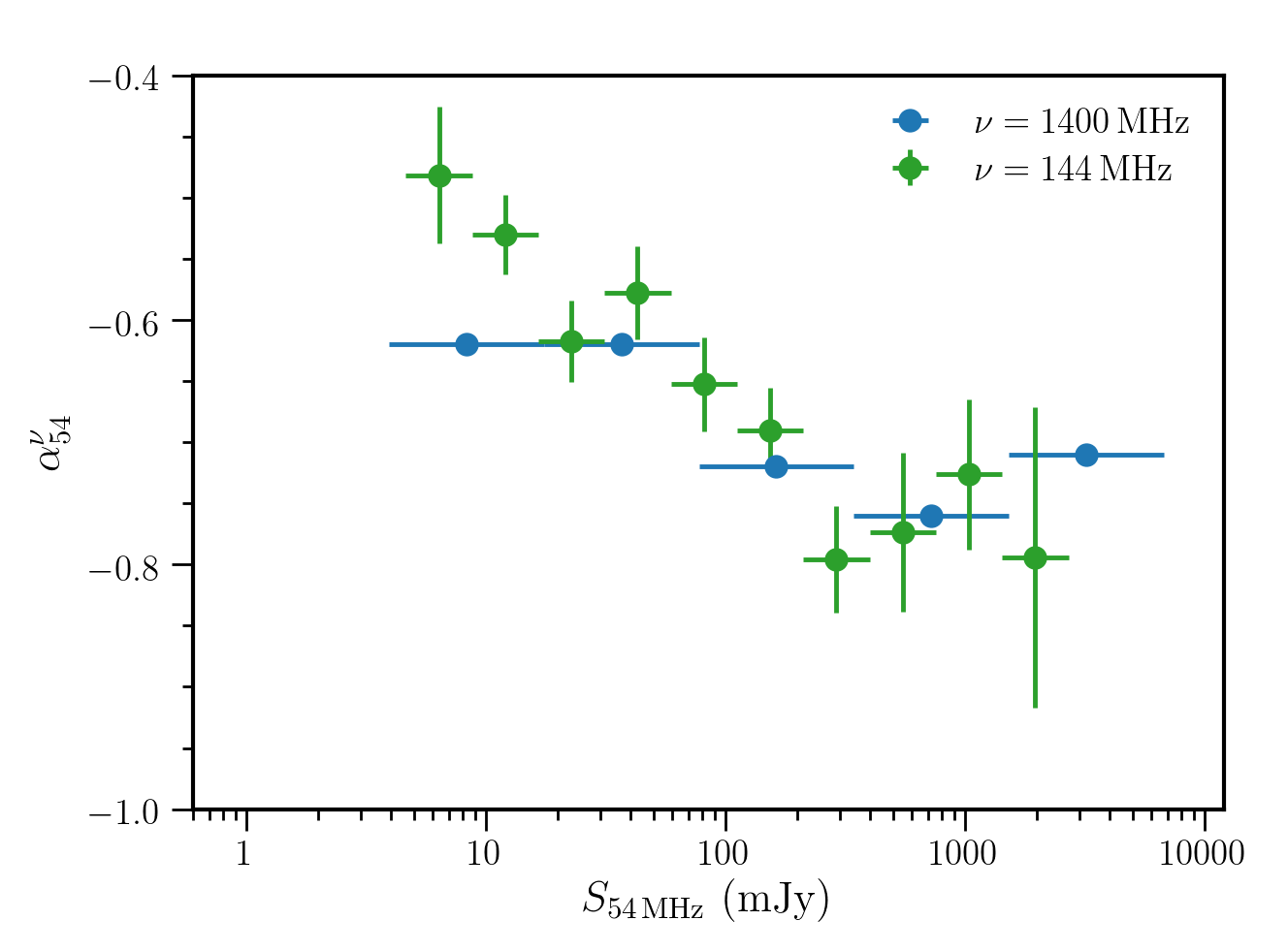}
\caption{Average spectral index of all sources between $1.4$\,GHz and $54$\,MHz determined from the source counts (in blue), and  between $144$\,MHz and $54$\,MHz determined from individual sources (from Fig.\,\ref{fig:specind}, in green).}
\label{fig:alpha_dep}
\end{figure}

In five flux density bins we determined the spectral index that provides the best scaling between the $1.4$-GHz and $54$-MHz source counts. This is plotted in Fig.\,\ref{fig:alpha_dep}, where we include the flux-density-dependent spectral index from  Fig.\,\ref{fig:specind}. Again this shows the flattening of the spectral index towards lower flux densities{, but in a statistical way, for the radio source populations  at these two frequencies\footnote{Individual matching to the 1.4-GHz data is deferred to a later work.}. This flattening is consistent with that observed between 54\,MHz and 144\,MHz for sources matched individually.} There is some indication that at $S_{54 \mathrm{MHz}}\sim10$\,\mJy{} the radio spectra at $54$--$144$\,MHz are flatter than at {$54$--$1400$\,MHz} suggesting curvature in the spectra of these sources. However, this may be due to the residual incompleteness in the $54$-MHz source counts and warrants further investigation with deeper LBA observations{, and detailed follow-up multi-frequency studies of the spectra of individual sources.}

\section{Summary}
\label{sect:concl}

We present the first deep ($\approx0.7$\,{\mJybeam}), high-resolution ($15 \times 15${\arcsec}) LOFAR LBA image of the Bo\"otes field made at $34$--$75$\,MHz from $56$ hours of observation and describe the full data reduction process from observations to  the direction-dependent calibrated image. The radio source catalogue of $1,289$ sources over an area of $23.6$\,\sqdeg{} allows us to characterise the low-frequency radio source population with unprecedented sensitivity. We present the Euclidean-normalised source counts and investigate the spectral indices of the source population, both of which indicate a flattening of the low-frequency radio spectra of fainter radio sources.

Our observations show that sub-mJy noise levels are obtainable with deep observations at these frequencies.  Additional observations of this field and other fields as part of LoLSS-Deep {will} allow us to probe even deeper, and to increase the area covered to provide better {source} statistics. Since the time of the observations presented here, a number of improvements have been made in the LOFAR LBA station calibration, resulting in improved sensitivity.  Further improvements in calibration are underway and will improve the imaging quality, particularly the dynamic range limitations around bright sources and the residual ionospheric smearing.

The combination of the very deep HBA data available for the Bo\"otes deep field and the wealth of multi-wavelength data, including redshifts and source classifications, will allow further detailed studies of the spectral properties of both star-forming galaxies and AGN.


\begin{acknowledgements}

The authors thank the anonymous referee whose feedback improved the paper and H.\,Edler for helpful discussions on ionospheric refraction.

WLW acknowledges support from the CAS-NWO programme for radio astronomy with project number 629.001.024, which is financed by the Netherlands Organisation for Scientific Research (NWO).
PNB is grateful for support from the UK STFC via grants ST/R000972/1 and ST/V000594/1.  MB acknowledges support from INAF under the SKA/CTA PRIN ``FORECaST'' and the PRIN MAIN STREAM ``SAuROS'' projects and from the Ministero degli Affari Esteri e della Cooperazione Internazionale - Direzione Generale per la Promozione del Sistema Paese Progetto di Grande Rilevanza ZA18GR02.


This research made use of \textsc{Astropy}, a community-developed core \textsc{python}
package for astronomy (Astropy Collaboration et al. 2013) hosted
at \url{http://www.astropy.org/}, and of \textsc{aplpy}, an open-source plotting
package for  \textsc{python} (Robitaille \& Bressert 2012).

LOFAR designed and constructed by ASTRON has facilities
in several countries, which are owned by various parties (each
with their own funding sources), and are collectively operated
by the International LOFAR Telescope (ILT) foundation under a
joint scientific policy. The ILT resources have benefited from the
following recent major funding sources: CNRS-INSU, Observatoire
de Paris and Universite d'Orl\'{e}ans, France; BMBF, MIWF-NRW, 
MPG, Germany; Science Foundation Ireland (SFI), Department of
Business, Enterprise and Innovation (DBEI), Ireland; NWO, the
Netherlands; the Science and Technology Facilities Council, UK;
and Ministry of Science and Higher Education, Poland; Istituto
Nazionale di Astrofisica (INAF). This research has made use of the University of
Hertfordshire high-performance computing facility (\url{https://uhhpc.herts.ac.uk/}) and the LOFAR-UK compute facility, located at the University of
Hertfordshire and supported by STFC [ST/P000096/1].

\end{acknowledgements}

\bibliographystyle{aa} 
\bibliography{deep_lba} 

\newcommand{\noop}[1]{}
\begin{thebibliography}{64}
\expandafter\ifx\csname natexlab\endcsname\relax\def\natexlab#1{#1}\fi

\bibitem[{{Baars} {et~al.}(1977){Baars}, {Genzel}, {Pauliny-Toth}, \&
  {Witzel}}]{Baars1977}
{Baars}, J.~W.~M., {Genzel}, R., {Pauliny-Toth}, I.~I.~K., \& {Witzel}, A.
  1977, \aap, 500, 135

\bibitem[{{Becker} {et~al.}(1995){Becker}, {White}, \& {Helfand}}]{Becker1995}
{Becker}, R.~H., {White}, R.~L., \& {Helfand}, D.~J. 1995, \apj, 450, 559

\bibitem[{{Belikov} {et~al.}(2011){Belikov}, {Dijkstra}, {Gankema}, {van Hoof},
  \& {Koopman}}]{Belikov2011}
{Belikov}, A.~N., {Dijkstra}, F., {Gankema}, J.~A., {van Hoof}, J.~B.~A.~N., \&
  {Koopman}, R. 2011, arXiv e-prints, arXiv:1110.0937

\bibitem[{{Bonato} {et~al.}(2017){Bonato}, {Negrello}, {Mancuso}, {De Zotti},
  {Ciliegi}, {Cai}, {Lapi}, {Massardi}, {Bonaldi}, {Sajina},
  {Smol{\v{c}}i{\'c}}, \& {Schinnerer}}]{Bonato2017}
{Bonato}, M., {Negrello}, M., {Mancuso}, C., {et~al.} 2017, \mnras, 469, 1912

\bibitem[{{Bonato} {et~al.}(2021){Bonato}, {Prandoni}, {De Zotti}, {Brienza},
  {Morganti}, \& {Vaccari}}]{Bonato2021}
{Bonato}, M., {Prandoni}, I., {De Zotti}, G., {et~al.} 2021, \mnras, 500, 22

\bibitem[{{Botteon} {et~al.}(2020){Botteon}, {van Weeren}, {Brunetti}, {de
  Gasperin}, {Intema}, {Osinga}, {Di Gennaro}, {Shimwell}, {Bonafede},
  {Br{\"u}ggen}, {Cassano}, {Cuciti}, {Dallacasa}, {Gastaldello}, {Mandal},
  {Rossetti}, \& {R{\"o}ttgering}}]{Botteon2020}
{Botteon}, A., {van Weeren}, R.~J., {Brunetti}, G., {et~al.} 2020, \mnras, 499,
  L11

\bibitem[{{Calistro Rivera} {et~al.}(2017){Calistro Rivera}, {Williams},
  {Hardcastle}, {Duncan}, {R{\"o}ttgering}, {Best}, {Br{\"u}ggen}, {Chy{\.z}y},
  {Conselice}, {de Gasperin}, {Engels}, {G{\"u}rkan}, {Intema}, {Jarvis},
  {Mahony}, {Miley}, {Morabito}, {Prandoni}, {Sabater}, {Smith}, {Tasse}, {van
  der Werf}, \& {White}}]{CalistroRivera2017}
{Calistro Rivera}, G., {Williams}, W.~L., {Hardcastle}, M.~J., {et~al.} 2017,
  \mnras, 469, 3468

\bibitem[{{Chambers} {et~al.}(2016){Chambers}, {Magnier}, {Metcalfe},
  {Flewelling}, {Huber}, {Waters}, {Denneau}, {Draper}, {Farrow}, {Finkbeiner},
  {Holmberg}, {Koppenhoefer}, {Price}, {Rest}, {Saglia}, {Schlafly}, {Smartt},
  {Sweeney}, {Wainscoat}, {Burgett}, {Chastel}, {Grav}, {Heasley}, {Hodapp},
  {Jedicke}, {Kaiser}, {Kudritzki}, {Luppino}, {Lupton}, {Monet}, {Morgan},
  {Onaka}, {Shiao}, {Stubbs}, {Tonry}, {White}, {Ba{\~n}ados}, {Bell},
  {Bender}, {Bernard}, {Boegner}, {Boffi}, {Botticella}, {Calamida},
  {Casertano}, {Chen}, {Chen}, {Cole}, {Deacon}, {Frenk}, {Fitzsimmons},
  {Gezari}, {Gibbs}, {Goessl}, {Goggia}, {Gourgue}, {Goldman}, {Grant},
  {Grebel}, {Hambly}, {Hasinger}, {Heavens}, {Heckman}, {Henderson}, {Henning},
  {Holman}, {Hopp}, {Ip}, {Isani}, {Jackson}, {Keyes}, {Koekemoer}, {Kotak},
  {Le}, {Liska}, {Long}, {Lucey}, {Liu}, {Martin}, {Masci}, {McLean}, {Mindel},
  {Misra}, {Morganson}, {Murphy}, {Obaika}, {Narayan}, {Nieto-Santisteban},
  {Norberg}, {Peacock}, {Pier}, {Postman}, {Primak}, {Rae}, {Rai}, {Riess},
  {Riffeser}, {Rix}, {R{\"o}ser}, {Russel}, {Rutz}, {Schilbach}, {Schultz},
  {Scolnic}, {Strolger}, {Szalay}, {Seitz}, {Small}, {Smith}, {Soderblom},
  {Taylor}, {Thomson}, {Taylor}, {Thakar}, {Thiel}, {Thilker}, {Unger},
  {Urata}, {Valenti}, {Wagner}, {Walder}, {Walter}, {Watters}, {Werner},
  {Wood-Vasey}, \& {Wyse}}]{Chambers2016}
{Chambers}, K.~C., {Magnier}, E.~A., {Metcalfe}, N., {et~al.} 2016, arXiv
  e-prints, arXiv:1612.05560

\bibitem[{{Condon}(1997)}]{Condon1997}
{Condon}, J.~J. 1997, \pasp, 109, 166

\bibitem[{{Coppejans} {et~al.}(2016){Coppejans}, {Cseh}, {van Velzen},
  {Falcke}, {Intema}, {Paragi}, {M{\"u}ller}, {Williams}, {Frey}, {Gurvits}, \&
  {K{\"o}rding}}]{Coppejans2016}
{Coppejans}, R., {Cseh}, D., {van Velzen}, S., {et~al.} 2016, \mnras, 459, 2455

\bibitem[{{Coppejans} {et~al.}(2015){Coppejans}, {Cseh}, {Williams}, {van
  Velzen}, \& {Falcke}}]{Coppejans2015}
{Coppejans}, R., {Cseh}, D., {Williams}, W.~L., {van Velzen}, S., \& {Falcke},
  H. 2015, \mnras, 450, 1477

\bibitem[{{Croft} {et~al.}(2008){Croft}, {van Breugel}, {Brown}, {de Vries},
  {Dey}, {Eisenhardt}, {Jannuzi}, {R{\"o}ttgering}, {Stanford}, {Stern}, \&
  {Willner}}]{Croft2008}
{Croft}, S., {van Breugel}, W., {Brown}, M.~J.~I., {et~al.} 2008, \aj, 135,
  1793

\bibitem[{{de Gasperin} {et~al.}(2020{\natexlab{a}}){de Gasperin}, {Brunetti},
  {Br{\"u}ggen}, {van Weeren}, {Williams}, {Botteon}, {Cuciti}, {Dijkema},
  {Edler}, {Iacobelli}, {Kang}, {Offringa}, {Orr{\'u}}, {Pizzo}, {Rafferty},
  {R{\"o}ttgering}, \& {Shimwell}}]{deGasperin2020a}
{de Gasperin}, F., {Brunetti}, G., {Br{\"u}ggen}, M., {et~al.}
  2020{\natexlab{a}}, \aap, 642, A85

\bibitem[{{de Gasperin} {et~al.}(2019){de Gasperin}, {Dijkema}, {Drabent},
  {Mevius}, {Rafferty}, {van Weeren}, {Br{\"u}ggen}, {Callingham}, {Emig},
  {Heald}, {Intema}, {Morabito}, {Offringa}, {Oonk}, {Orr{\`u}},
  {R{\"o}ttgering}, {Sabater}, {Shimwell}, {Shulevski}, \&
  {Williams}}]{deGasperin2019}
{de Gasperin}, F., {Dijkema}, T.~J., {Drabent}, A., {et~al.} 2019, \aap, 622,
  A5

\bibitem[{{de Gasperin} {et~al.}(2018{\natexlab{a}}){de Gasperin}, {Intema}, \&
  {Frail}}]{deGasperin2018a}
{de Gasperin}, F., {Intema}, H.~T., \& {Frail}, D.~A. 2018{\natexlab{a}},
  \mnras, 474, 5008

\bibitem[{{de Gasperin} {et~al.}(2020{\natexlab{b}}){de Gasperin}, {Lazio}, \&
  {Knapp}}]{deGasperin2020b}
{de Gasperin}, F., {Lazio}, T.~J.~W., \& {Knapp}, M. 2020{\natexlab{b}}, \aap,
  644, A157

\bibitem[{{de Gasperin} {et~al.}(2018{\natexlab{b}}){de Gasperin}, {Mevius},
  {Rafferty}, {Intema}, \& {Fallows}}]{deGasperin2018b}
{de Gasperin}, F., {Mevius}, M., {Rafferty}, D.~A., {Intema}, H.~T., \&
  {Fallows}, R.~A. 2018{\natexlab{b}}, \aap, 615, A179

\bibitem[{{de Gasperin} {et~al.}(2012){de Gasperin}, {Orr{\'u}}, {Murgia},
  {Merloni}, {Falcke}, {Beck}, {Beswick}, {B{\^\i}rzan}, {Bonafede},
  {Br{\"u}ggen}, {Brunetti}, {Chy{\.z}y}, {Conway}, {Croston}, {En{\ss}lin},
  {Ferrari}, {Heald}, {Heidenreich}, {Jackson}, {Macario}, {McKean}, {Miley},
  {Morganti}, {Offringa}, {Pizzo}, {Rafferty}, {R{\"o}ttgering}, {Shulevski},
  {Steinmetz}, {Tasse}, {van der Tol}, {van Driel}, {van Weeren}, {van
  Zwieten}, {Alexov}, {Anderson}, {Asgekar}, {Avruch}, {Bell}, {Bell},
  {Bentum}, {Bernardi}, {Best}, {Breitling}, {Broderick}, {Butcher}, {Ciardi},
  {Dettmar}, {Eisloeffel}, {Frieswijk}, {Gankema}, {Garrett}, {Gerbers},
  {Griessmeier}, {Gunst}, {Hassall}, {Hessels}, {Hoeft}, {Horneffer},
  {Karastergiou}, {K{\"o}hler}, {Koopman}, {Kuniyoshi}, {Kuper}, {Maat},
  {Mann}, {Mevius}, {Mulcahy}, {Munk}, {Nijboer}, {Noordam}, {Paas}, {Pandey},
  {Pandey}, {Polatidis}, {Reich}, {Schoenmakers}, {Sluman}, {Smirnov}, {Sobey},
  {Stappers}, {Swinbank}, {Tagger}, {Tang}, {van Bemmel}, {van Cappellen}, {van
  Duin}, {van Haarlem}, {van Leeuwen}, {Vermeulen}, {Vocks}, {White}, {Wise},
  {Wucknitz}, \& {Zarka}}]{deGasperin2012}
{de Gasperin}, F., {Orr{\'u}}, E., {Murgia}, M., {et~al.} 2012, \aap, 547, A56

\bibitem[{{de Gasperin} {et~al.}(2021){de Gasperin}, {Williams}, {Best},
  {Br{\"u}ggen}, {Brunetti}, {Cuciti}, {Dijkema}, {Hardcastle}, {Norden},
  {Offringa}, {Shimwell}, {van Weeren}, {Bomans}, {Bonafede}, {Botteon},
  {Callingham}, {Cassano}, {Chy{\.z}y}, {Emig}, {Edler}, {Haverkorn}, {Heald},
  {Heesen}, {Iacobelli}, {Intema}, {Kadler}, {Ma{\l}ek}, {Mevius}, {Miley},
  {Mingo}, {Morabito}, {Sabater}, {Morganti}, {Orr{\'u}}, {Pizzo}, {Prandoni},
  {Shulevski}, {Tasse}, {Vaccari}, {Zarka}, \&
  {R{\"o}ttgering}}]{deGasperin2021}
{de Gasperin}, F., {Williams}, W.~L., {Best}, P., {et~al.} 2021, \aap, 648,
  A104

\bibitem[{{de Vries} {et~al.}(2002){de Vries}, {Morganti}, {R{\"o}ttgering},
  {Vermeulen}, {van Breugel}, {Rengelink}, \& {Jarvis}}]{deVries2002}
{de Vries}, W.~H., {Morganti}, R., {R{\"o}ttgering}, H.~J.~A., {et~al.} 2002,
  \aj, 123, 1784

\bibitem[{{de Zotti} {et~al.}(2010){de Zotti}, {Massardi}, {Negrello}, \&
  {Wall}}]{deZotti2010}
{de Zotti}, G., {Massardi}, M., {Negrello}, M., \& {Wall}, J. 2010, \aapr, 18,
  1

\bibitem[{{Duncan} {et~al.}(2021){Duncan}, {Kondapally}, {Brown}, {Bonato},
  {Best}, {R{\"o}ttgering}, {Bondi}, {Bowler}, {Cochrane}, {G{\"u}rkan},
  {Hardcastle}, {Jarvis}, {Kunert-Bajraszewska}, {Leslie}, {Ma{\l}ek},
  {Morabito}, {O'Sullivan}, {Prandoni}, {Sabater}, {Shimwell}, {Smith}, {Wang},
  {Wo{\l}owska}, \& {Tasse}}]{Duncan2021}
{Duncan}, K.~J., {Kondapally}, R., {Brown}, M.~J.~I., {et~al.} 2021, \aap, 648,
  A4

\bibitem[{{Eisenhardt} {et~al.}(2004){Eisenhardt}, {Stern}, {Brodwin}, {Fazio},
  {Rieke}, {Rieke}, {Werner}, {Wright}, {Allen}, {Arendt}, {Ashby}, {Barmby},
  {Forrest}, {Hora}, {Huang}, {Huchra}, {Pahre}, {Pipher}, {Reach}, {Smith},
  {Stauffer}, {Wang}, {Willner}, {Brown}, {Dey}, {Jannuzi}, \&
  {Tiede}}]{Eisenhardt2004}
{Eisenhardt}, P.~R., {Stern}, D., {Brodwin}, M., {et~al.} 2004, \apjs, 154, 48

\bibitem[{{Gehrels}(1986)}]{Gehrels1986}
{Gehrels}, N. 1986, \apj, 303, 336

\bibitem[{{Hales} {et~al.}(1988){Hales}, {Baldwin}, \& {Warner}}]{Hales1988}
{Hales}, S.~E.~G., {Baldwin}, J.~E., \& {Warner}, P.~J. 1988, \mnras, 234, 919

\bibitem[{{Higdon} {et~al.}(2005){Higdon}, {Higdon}, {Weedman}, {Houck}, {Le
  Floc'h}, {Brown}, {Dey}, {Jannuzi}, {Soifer}, \& {Rieke}}]{Higdon2005}
{Higdon}, J.~L., {Higdon}, S.~J.~U., {Weedman}, D.~W., {et~al.} 2005, \apj,
  626, 58

\bibitem[{{Jannuzi} {et~al.}(1999){Jannuzi}, {Dey}, \& {NDWFS
  Team}}]{Jannuzi1999}
{Jannuzi}, B.~T., {Dey}, A., \& {NDWFS Team}. 1999, in Bulletin of the American
  Astronomical Society, Vol.~31, American Astronomical Society Meeting
  Abstracts, 1392

\bibitem[{{Kellermann} {et~al.}(1969){Kellermann}, {Pauliny-Toth}, \&
  {Williams}}]{KPW1969}
{Kellermann}, K.~I., {Pauliny-Toth}, I.~I.~K., \& {Williams}, P.~J.~S. 1969,
  \apj, 157, 1

\bibitem[{{Kenter} {et~al.}(2005){Kenter}, {Murray}, {Forman}, {Jones},
  {Green}, {Kochanek}, {Vikhlinin}, {Fabricant}, {Fazio}, {Brand}, {Brown},
  {Dey}, {Jannuzi}, {Najita}, {McNamara}, {Shields}, \& {Rieke}}]{Kenter2005}
{Kenter}, A., {Murray}, S.~S., {Forman}, W.~R., {et~al.} 2005, \apjs, 161, 9

\bibitem[{{Kochanek} {et~al.}(2012){Kochanek}, {Eisenstein}, {Cool},
  {Caldwell}, {Assef}, {Jannuzi}, {Jones}, {Murray}, {Forman}, {Dey}, {Brown},
  {Eisenhardt}, {Gonzalez}, {Green}, \& {Stern}}]{Kochanek2012}
{Kochanek}, C.~S., {Eisenstein}, D.~J., {Cool}, R.~J., {et~al.} 2012, \apjs,
  200, 8

\bibitem[{{Kondapally} {et~al.}(2021){Kondapally}, {Best}, {Hardcastle},
  {Nisbet}, {Bonato}, {Sabater}, {Duncan}, {McCheyne}, {Cochrane}, {Bowler},
  {Williams}, {Shimwell}, {Tasse}, {Croston}, {Goyal}, {Jamrozy}, {Jarvis},
  {Mahatma}, {R{\"o}ttgering}, {Smith}, {Wo{\l}owska}, {Bondi}, {Brienza},
  {Brown}, {Br{\"u}ggen}, {Chambers}, {Garrett}, {G{\"u}rkan}, {Huber},
  {Kunert-Bajraszewska}, {Magnier}, {Mingo}, {Mostert},
  {Nikiel-Wroczy{\'n}ski}, {O'Sullivan}, {Paladino}, {Ploeckinger}, {Prandoni},
  {Rosenthal}, {Schwarz}, {Shulevski}, {Wagenveld}, \& {Wang}}]{Kondapally2021}
{Kondapally}, R., {Best}, P.~N., {Hardcastle}, M.~J., {et~al.} 2021, \aap, 648,
  A3

\bibitem[{{Lane} {et~al.}(2014){Lane}, {Cotton}, {van Velzen}, {Clarke},
  {Kassim}, {Helmboldt}, {Lazio}, \& {Cohen}}]{Lane2014}
{Lane}, W.~M., {Cotton}, W.~D., {van Velzen}, S., {et~al.} 2014, \mnras, 440,
  327

\bibitem[{{Mandal} {et~al.}(2021){Mandal}, {Prandoni}, {Hardcastle},
  {Shimwell}, {Intema}, {Tasse}, {van Weeren}, {Algera}, {Emig},
  {R{\"o}ttgering}, {Schwarz}, {Siewert}, {Best}, {Bonato}, {Bondi}, {Jarvis},
  {Kondapally}, {Leslie}, {Mahatma}, {Sabater}, {Retana-Montenegro}, \&
  {Williams}}]{Mandal2021}
{Mandal}, S., {Prandoni}, I., {Hardcastle}, M.~J., {et~al.} 2021, \aap, 648, A5

\bibitem[{{Martin} {et~al.}(2003){Martin}, {Barlow}, {Barnhart}, {Bianchi},
  {Blakkolb}, {Bruno}, {Bushman}, {Byun}, {Chiville}, {Conrow}, {Cooke},
  {Donas}, {Fanson}, {Forster}, {Friedman}, {Grange}, {Griffiths}, {Heckman},
  {Lee}, {Jelinsky}, {Kim}, {Lee}, {Lee}, {Liu}, {Madore}, {Malina}, {Mazer},
  {McLean}, {Milliard}, {Mitchell}, {Morais}, {Morrissey}, {Neff}, {Raison},
  {Randall}, {Rich}, {Schiminovich}, {Schmitigal}, {Sen}, {Siegmund}, {Small},
  {Stock}, {Surber}, {Szalay}, {Vaughan}, {Weigand}, {Welsh}, {Wu}, {Wyder},
  {Xu}, \& {Zsoldas}}]{Martin2003}
{Martin}, C., {Barlow}, T., {Barnhart}, W., {et~al.} 2003, in Society of
  Photo-Optical Instrumentation Engineers (SPIE) Conference Series, Vol. 4854,
  Future EUV/UV and Visible Space Astrophysics Missions and Instrumentation.,
  ed. J.~C. {Blades} \& O.~H.~W. {Siegmund}, 336--350

\bibitem[{{McConnell} {et~al.}(2020){McConnell}, {Hale}, {Lenc}, {Banfield},
  {Heald}, {Hotan}, {Leung}, {Moss}, {Murphy}, {O'Brien}, {Pritchard}, {Raja},
  {Sadler}, {Stewart}, {Thomson}, {Whiting}, {Allison}, {Amy}, {Anderson},
  {Ball}, {Bannister}, {Bell}, {Bock}, {Bolton}, {Bunton}, {Chippendale},
  {Collier}, {Cooray}, {Cornwell}, {Diamond}, {Edwards}, {Gupta}, {Hayman},
  {Heywood}, {Jackson}, {Koribalski}, {Lee-Waddell}, {McClure-Griffiths}, {Ng},
  {Norris}, {Phillips}, {Reynolds}, {Roxby}, {Schinckel}, {Shields},
  {Tremblay}, {Tzioumis}, {Voronkov}, \& {Westmeier}}]{McConnell2020}
{McConnell}, D., {Hale}, C.~L., {Lenc}, E., {et~al.} 2020, \pasa, 37, e048

\bibitem[{{Mevius} {et~al.}(2016){Mevius}, {van der Tol}, {Pandey},
  {Vedantham}, {Brentjens}, {de Bruyn}, {Abdalla}, {Asad}, {Bregman}, {Brouw},
  {Bus}, {Chapman}, {Ciardi}, {Fernandez}, {Ghosh}, {Harker}, {Iliev},
  {Jeli{\'c}}, {Kazemi}, {Koopmans}, {Noordam}, {Offringa}, {Patil}, {van
  Weeren}, {Wijnholds}, {Yatawatta}, \& {Zaroubi}}]{Mevius2016}
{Mevius}, M., {van der Tol}, S., {Pandey}, V.~N., {et~al.} 2016, Radio Science,
  51, 927

\bibitem[{{Mohan} \& {Rafferty}(2015)}]{Mohan2015}
{Mohan}, N. \& {Rafferty}, D. 2015, {PyBDSF: Python Blob Detection and Source
  Finder}

\bibitem[{{Murray} {et~al.}(2005){Murray}, {Kenter}, {Forman}, {Jones},
  {Green}, {Kochanek}, {Vikhlinin}, {Fabricant}, {Fazio}, {Brand}, {Brown},
  {Dey}, {Jannuzi}, {Najita}, {McNamara}, {Shields}, \& {Rieke}}]{Murray2005}
{Murray}, S.~S., {Kenter}, A., {Forman}, W.~R., {et~al.} 2005, \apjs, 161, 1

\bibitem[{{Offringa}(2010)}]{Offringa2010}
{Offringa}, A.~R. 2010, {AOFlagger: RFI Software}

\bibitem[{{Offringa}(2016)}]{Offringa2016}
{Offringa}, A.~R. 2016, \aap, 595, A99

\bibitem[{Offringa {et~al.}(2014)Offringa, McKinley, Hurley-Walker,
  {et~al.}}]{Offringa2014}
Offringa, A.~R., McKinley, B., Hurley-Walker, {et~al.} 2014, MNRAS, 444, 606

\bibitem[{{Offringa} {et~al.}(2012){Offringa}, {van de Gronde}, \&
  {Roerdink}}]{Offringa2012}
{Offringa}, A.~R., {van de Gronde}, J.~J., \& {Roerdink}, J.~B.~T.~M. 2012,
  \aap, 539, A95

\bibitem[{{Prandoni} {et~al.}(2001){Prandoni}, {Gregorini}, {Parma}, {de
  Ruiter}, {Vettolani}, {Wieringa}, \& {Ekers}}]{Prandoni2001}
{Prandoni}, I., {Gregorini}, L., {Parma}, P., {et~al.} 2001, \aap, 365, 392

\bibitem[{{Ramasawmy} {et~al.}(2021){Ramasawmy}, {Geach}, {Hardcastle}, {Best},
  {Bonato}, {Bondi}, {Calistro Rivera}, {Cochrane}, {Conway}, {Coppin},
  {Duncan}, {Dunlop}, {Franco}, {Garc{\'\i}a-Vergara}, {Jarvis}, {Kondapally},
  {McCheyne}, {Prandoni}, {R{\"o}ttgering}, {Smith}, {Tasse}, \&
  {Wang}}]{Ramasawmy2021}
{Ramasawmy}, J., {Geach}, J.~E., {Hardcastle}, M.~J., {et~al.} 2021, \aap, 648,
  A14

\bibitem[{{Rengelink} {et~al.}(1997){Rengelink}, {Tang}, {de Bruyn}, {Miley},
  {Bremer}, {Roettgering}, \& {Bremer}}]{Rengelink1997}
{Rengelink}, R.~B., {Tang}, Y., {de Bruyn}, A.~G., {et~al.} 1997, \aaps, 124,
  259

\bibitem[{Reynolds(1994)}]{Reynolds1994}
Reynolds, J. 1994, A revised flux scale for the AT Compact Array, Tech. Rep.
  AT/39.3/040, ATNF

\bibitem[{{Roger} {et~al.}(1973){Roger}, {Costain}, \& {Bridle}}]{RCB1973}
{Roger}, R.~S., {Costain}, C.~H., \& {Bridle}, A.~H. 1973, \aj, 78, 1030

\bibitem[{{Sabater} {et~al.}(2021){Sabater}, {Best}, {Tasse}, {Hardcastle},
  {Shimwell}, {Nisbet}, {Jelic}, {Callingham}, {R{\"o}ttgering}, {Bonato},
  {Bondi}, {Ciardi}, {Cochrane}, {Jarvis}, {Kondapally}, {Koopmans},
  {O'Sullivan}, {Prandoni}, {Schwarz}, {Smith}, {Wang}, {Williams}, \&
  {Zaroubi}}]{Sabater2021}
{Sabater}, J., {Best}, P.~N., {Tasse}, C., {et~al.} 2021, \aap, 648, A2

\bibitem[{{Scaife} \& {Heald}(2012)}]{ScaifeHeald2012}
{Scaife}, A. M.~M. \& {Heald}, G.~H. 2012, \mnras, 423, L30

\bibitem[{{Shimwell} {et~al.}(2017){Shimwell}, {R{\"o}ttgering}, {Best},
  {Williams}, {Dijkema}, {de Gasperin}, {Hardcastle}, {Heald}, {Hoang},
  {Horneffer}, {Intema}, {Mahony}, {Mandal}, {Mechev}, {Morabito}, {Oonk},
  {Rafferty}, {Retana-Montenegro}, {Sabater}, {Tasse}, {van Weeren},
  {Br{\"u}ggen}, {Brunetti}, {Chy{\.z}y}, {Conway}, {Haverkorn}, {Jackson},
  {Jarvis}, {McKean}, {Miley}, {Morganti}, {White}, {Wise}, {van Bemmel},
  {Beck}, {Brienza}, {Bonafede}, {Calistro Rivera}, {Cassano}, {Clarke},
  {Cseh}, {Deller}, {Drabent}, {van Driel}, {Engels}, {Falcke}, {Ferrari},
  {Fr{\"o}hlich}, {Garrett}, {Harwood}, {Heesen}, {Hoeft}, {Horellou},
  {Israel}, {Kapi{\'n}ska}, {Kunert-Bajraszewska}, {McKay}, {Mohan},
  {Orr{\'u}}, {Pizzo}, {Prandoni}, {Schwarz}, {Shulevski}, {Sipior}, {Smith},
  {Sridhar}, {Steinmetz}, {Stroe}, {Varenius}, {van der Werf}, {Zensus}, \&
  {Zwart}}]{Shimwell2017}
{Shimwell}, T.~W., {R{\"o}ttgering}, H.~J.~A., {Best}, P.~N., {et~al.} 2017,
  \aap, 598, A104

\bibitem[{{Shimwell} {et~al.}(2019){Shimwell}, {Tasse}, {Hardcastle}, {Mechev},
  {Williams}, {Best}, {R{\"o}ttgering}, {Callingham}, {Dijkema}, {de Gasperin},
  {Hoang}, {Hugo}, {Mirmont}, {Oonk}, {Prandoni}, {Rafferty}, {Sabater},
  {Smirnov}, {van Weeren}, {White}, {Atemkeng}, {Bester}, {Bonnassieux},
  {Br{\"u}ggen}, {Brunetti}, {Chy{\.z}y}, {Cochrane}, {Conway}, {Croston},
  {Danezi}, {Duncan}, {Haverkorn}, {Heald}, {Iacobelli}, {Intema}, {Jackson},
  {Jamrozy}, {Jarvis}, {Lakhoo}, {Mevius}, {Miley}, {Morabito}, {Morganti},
  {Nisbet}, {Orr{\'u}}, {Perkins}, {Pizzo}, {Schrijvers}, {Smith}, {Vermeulen},
  {Wise}, {Alegre}, {Bacon}, {van Bemmel}, {Beswick}, {Bonafede}, {Botteon},
  {Bourke}, {Brienza}, {Calistro Rivera}, {Cassano}, {Clarke}, {Conselice},
  {Dettmar}, {Drabent}, {Dumba}, {Emig}, {En{\ss}lin}, {Ferrari}, {Garrett},
  {G{\'e}nova-Santos}, {Goyal}, {G{\"u}rkan}, {Hale}, {Harwood}, {Heesen},
  {Hoeft}, {Horellou}, {Jackson}, {Kokotanekov}, {Kondapally},
  {Kunert-Bajraszewska}, {Mahatma}, {Mahony}, {Mandal}, {McKean}, {Merloni},
  {Mingo}, {Miskolczi}, {Mooney}, {Nikiel-Wroczy{\'n}ski}, {O'Sullivan},
  {Quinn}, {Reich}, {Roskowi{\'n}ski}, {Rowlinson}, {Savini}, {Saxena},
  {Schwarz}, {Shulevski}, {Sridhar}, {Stacey}, {Urquhart}, {van der Wiel},
  {Varenius}, {Webster}, \& {Wilber}}]{Shimwell2019}
{Shimwell}, T.~W., {Tasse}, C., {Hardcastle}, M.~J., {et~al.} 2019, \aap, 622,
  A1

\bibitem[{{Smirnov} \& {Tasse}(2015)}]{SmirnovTasse2015}
{Smirnov}, O.~M. \& {Tasse}, C. 2015, \mnras, 449, 2668

\bibitem[{{Tasse}(2014{\natexlab{a}})}]{Tasse2014a}
{Tasse}, C. 2014{\natexlab{a}}, arXiv e-prints, arXiv:1410.8706

\bibitem[{{Tasse}(2014{\natexlab{b}})}]{Tasse2014b}
{Tasse}, C. 2014{\natexlab{b}}, \aap, 566, A127

\bibitem[{{Tasse} {et~al.}(2018){Tasse}, {Hugo}, {Mirmont}, {Smirnov},
  {Atemkeng}, {Bester}, {Hardcastle}, {Lakhoo}, {Perkins}, \&
  {Shimwell}}]{Tasse2018}
{Tasse}, C., {Hugo}, B., {Mirmont}, M., {et~al.} 2018, \aap, 611, A87

\bibitem[{{Tasse} {et~al.}(2021){Tasse}, {Shimwell}, {Hardcastle},
  {O'Sullivan}, {van Weeren}, {Best}, {Bester}, {Hugo}, {Smirnov}, {Sabater},
  {Calistro-Rivera}, {de Gasperin}, {Morabito}, {R{\"o}ttgering}, {Williams},
  {Bonato}, {Bondi}, {Botteon}, {Br{\"u}ggen}, {Brunetti}, {Chy{\.z}y},
  {Garrett}, {G{\"u}rkan}, {Jarvis}, {Kondapally}, {Mandal}, {Prandoni},
  {Repetti}, {Retana-Montenegro}, {Schwarz}, {Shulevski}, \&
  {Wiaux}}]{Tasse2021}
{Tasse}, C., {Shimwell}, T., {Hardcastle}, M.~J., {et~al.} 2021, \aap, 648, A1

\bibitem[{{van der Tol} {et~al.}(2007){van der Tol}, {Jeffs}, \& {van der
  Veen}}]{vanderTol2007}
{van der Tol}, S., {Jeffs}, B.~D., \& {van der Veen}, A.~J. 2007, IEEE
  Transactions on Signal Processing, 55, 4497

\bibitem[{{van Diepen} {et~al.}(2018){van Diepen}, {Dijkema}, \&
  {Offringa}}]{vanDiepen2018}
{van Diepen}, G., {Dijkema}, T.~J., \& {Offringa}, A. 2018, {DPPP: Default
  Pre-Processing Pipeline}

\bibitem[{{van Diepen}(2015)}]{vanDiepen2015}
{van Diepen}, G.~N.~J. 2015, Astronomy and Computing, 12, 174

\bibitem[{{van Haarlem} {et~al.}(2013){van Haarlem}, {Wise}, {Gunst}, {Heald},
  {McKean}, {Hessels}, {de Bruyn}, {Nijboer}, {Swinbank}, {Fallows},
  {Brentjens}, {Nelles}, {Beck}, {Falcke}, {Fender}, {H{\"o}randel},
  {Koopmans}, {Mann}, {Miley}, {R{\"o}ttgering}, {Stappers}, {Wijers},
  {Zaroubi}, {van den Akker}, {Alexov}, {Anderson}, {Anderson}, {van Ardenne},
  {Arts}, {Asgekar}, {Avruch}, {Batejat}, {B{\"a}hren}, {Bell}, {Bell}, {van
  Bemmel}, {Bennema}, {Bentum}, {Bernardi}, {Best}, {B{\^\i}rzan}, {Bonafede},
  {Boonstra}, {Braun}, {Bregman}, {Breitling}, {van de Brink}, {Broderick},
  {Broekema}, {Brouw}, {Br{\"u}ggen}, {Butcher}, {van Cappellen}, {Ciardi},
  {Coenen}, {Conway}, {Coolen}, {Corstanje}, {Damstra}, {Davies}, {Deller},
  {Dettmar}, {van Diepen}, {Dijkstra}, {Donker}, {Doorduin}, {Dromer}, {Drost},
  {van Duin}, {Eisl{\"o}ffel}, {van Enst}, {Ferrari}, {Frieswijk}, {Gankema},
  {Garrett}, {de Gasperin}, {Gerbers}, {de Geus}, {Grie{\ss}meier}, {Grit},
  {Gruppen}, {Hamaker}, {Hassall}, {Hoeft}, {Holties}, {Horneffer}, {van der
  Horst}, {van Houwelingen}, {Huijgen}, {Iacobelli}, {Intema}, {Jackson},
  {Jelic}, {de Jong}, {Juette}, {Kant}, {Karastergiou}, {Koers}, {Kollen},
  {Kondratiev}, {Kooistra}, {Koopman}, {Koster}, {Kuniyoshi}, {Kramer},
  {Kuper}, {Lambropoulos}, {Law}, {van Leeuwen}, {Lemaitre}, {Loose}, {Maat},
  {Macario}, {Markoff}, {Masters}, {McFadden}, {McKay-Bukowski}, {Meijering},
  {Meulman}, {Mevius}, {Middelberg}, {Millenaar}, {Miller-Jones}, {Mohan},
  {Mol}, {Morawietz}, {Morganti}, {Mulcahy}, {Mulder}, {Munk}, {Nieuwenhuis},
  {van Nieuwpoort}, {Noordam}, {Norden}, {Noutsos}, {Offringa}, {Olofsson},
  {Omar}, {Orr{\'u}}, {Overeem}, {Paas}, {Pand ey-Pommier}, {Pandey}, {Pizzo},
  {Polatidis}, {Rafferty}, {Rawlings}, {Reich}, {de Reijer}, {Reitsma},
  {Renting}, {Riemers}, {Rol}, {Romein}, {Roosjen}, {Ruiter}, {Scaife}, {van
  der Schaaf}, {Scheers}, {Schellart}, {Schoenmakers}, {Schoonderbeek},
  {Serylak}, {Shulevski}, {Sluman}, {Smirnov}, {Sobey}, {Spreeuw}, {Steinmetz},
  {Sterks}, {Stiepel}, {Stuurwold}, {Tagger}, {Tang}, {Tasse}, {Thomas},
  {Thoudam}, {Toribio}, {van der Tol}, {Usov}, {van Veelen}, {van der Veen},
  {ter Veen}, {Verbiest}, {Vermeulen}, {Vermaas}, {Vocks}, {Vogt}, {de Vos},
  {van der Wal}, {van Weeren}, {Weggemans}, {Weltevrede}, {White}, {Wijnholds},
  {Wilhelmsson}, {Wucknitz}, {Yatawatta}, {Zarka}, {Zensus}, \& {van
  Zwieten}}]{vanHaarlem2013}
{van Haarlem}, M.~P., {Wise}, M.~W., {Gunst}, A.~W., {et~al.} 2013, \aap, 556,
  A2

\bibitem[{{van Weeren} {et~al.}(2014){van Weeren}, {Williams}, {Tasse},
  {R{\"o}ttgering}, {Rafferty}, {van der Tol}, {Heald}, {White}, {Shulevski},
  {Best}, {Intema}, {Bhatnagar}, {Reich}, {Steinmetz}, {van Velzen},
  {En{\ss}lin}, {Prand oni}, {de Gasperin}, {Jamrozy}, {Brunetti}, {Jarvis},
  {McKean}, {Wise}, {Ferrari}, {Harwood}, {Oonk}, {Hoeft},
  {Kunert-Bajraszewska}, {Horellou}, {Wucknitz}, {Bonafede}, {Mohan}, {Scaife},
  {Kl{\"o}ckner}, {van Bemmel}, {Merloni}, {Chyzy}, {Engels}, {Falcke}, {Pand
  ey-Pommier}, {Alexov}, {Anderson}, {Avruch}, {Beck}, {Bell}, {Bentum},
  {Bernardi}, {Breitling}, {Broderick}, {Brouw}, {Br{\"u}ggen}, {Butcher},
  {Ciardi}, {de Geus}, {de Vos}, {Deller}, {Duscha}, {Eisl{\"o}ffel},
  {Fallows}, {Frieswijk}, {Garrett}, {Grie{\ss}meier}, {Gunst}, {Hamaker},
  {Hassall}, {H{\"o}randel}, {van der Horst}, {Iacobelli}, {Jackson}, {Juette},
  {Kondratiev}, {Kuniyoshi}, {Maat}, {Mann}, {McKay-Bukowski}, {Mevius},
  {Morganti}, {Munk}, {Offringa}, {Orr{\`u}}, {Paas}, {Pandey}, {Pietka},
  {Pizzo}, {Polatidis}, {Renting}, {Rowlinson}, {Schwarz}, {Serylak}, {Sluman},
  {Smirnov}, {Stappers}, {Stewart}, {Swinbank}, {Tagger}, {Tang}, {Thoudam},
  {Toribio}, {Vermeulen}, {Vocks}, \& {Zarka}}]{vanWeeren2014}
{van Weeren}, R.~J., {Williams}, W.~L., {Tasse}, C., {et~al.} 2014, \apj, 793,
  82

\bibitem[{{Williams} {et~al.}(2013){Williams}, {Intema}, \&
  {R{\"o}ttgering}}]{Williams2013}
{Williams}, W.~L., {Intema}, H.~T., \& {R{\"o}ttgering}, H.~J.~A. 2013, \aap,
  549, A55

\bibitem[{{Williams} {et~al.}(2016){Williams}, {van Weeren}, {R{\"o}ttgering},
  {Best}, {Dijkema}, {de Gasperin}, {Hardcastle}, {Heald}, {Prand oni},
  {Sabater}, {Shimwell}, {Tasse}, {van Bemmel}, {Br{\"u}ggen}, {Brunetti},
  {Conway}, {En{\ss}lin}, {Engels}, {Falcke}, {Ferrari}, {Haverkorn},
  {Jackson}, {Jarvis}, {Kapi{\'n}ska}, {Mahony}, {Miley}, {Morabito},
  {Morganti}, {Orr{\'u}}, {Retana-Montenegro}, {Sridhar}, {Toribio}, {White},
  {Wise}, \& {Zwart}}]{Williams2016}
{Williams}, W.~L., {van Weeren}, R.~J., {R{\"o}ttgering}, H.~J.~A., {et~al.}
  2016, \mnras, 460, 2385

\bibitem[{{Wilman} {et~al.}(2008){Wilman}, {Miller}, {Jarvis}, {Mauch},
  {Levrier}, {Abdalla}, {Rawlings}, {Kl{\"o}ckner}, {Obreschkow}, {Olteanu}, \&
  {Young}}]{Wilman2008}
{Wilman}, R.~J., {Miller}, L., {Jarvis}, M.~J., {et~al.} 2008, \mnras, 388,
  1335

\end{thebibliography}

\end{document}